\documentclass{aa}  

\usepackage{graphicx}
\usepackage{txfonts}
\usepackage{graphicx}
\usepackage{soul}
\usepackage{natbib}

\usepackage[colorlinks=true, citecolor=blue, urlcolor=blue, breaklinks=true, bookmarks=true, bookmarksopen=true]{hyperref}
\usepackage{subcaption}
\begin{document} 

    \title{Revealing the stellar population of the ultra-obscured Galactic globular cluster Glimpse-C02\thanks{Based on observations with the NASA/ESA HST, obtained under program GO 17918 (PI: Loriga). The Space Telescope Science Institute is operated by AURA, Inc., under NASA contract NAS5-26555.}}

   \author{M. Loriga
          \inst{1,2}
          \and
          M. Cadelano 
          \inst{1,2}
          \and
          C. Pallanca
          \inst{1,2}
          \and
          F. R. Ferraro 
          \inst{1,2}
          \and
          B. Lanzoni 
          \inst{1,2}
          \and
          L. Chiappino
          \inst{1,2}
          \and
          C. Crociati
          \inst{3}
          \and
          E. Dalessandro 
          \inst{2}
          \and
          C. Giusti 
          \inst{1,2}
          \and
          S. Leanza
          \inst{1,2}
          \and
          D. Massari
          \inst{2}
          \and
          L. Origlia
          \inst{2}
          \and
          E. Vesperini
          \inst{4}
          }

   \institute{Dipartimento di Fisica e Astronomia, Università degli Studi di Bologna,
              Via Piero Gobetti 93/2, I-40129 Bologna
         \and
             INAF - Astrophysics and Space Science Observatory Bologna , 
             Via Piero Gobetti 93/3, I-40129 Bologna
         \and
            Institute for Astronomy, University of Edinburgh, Royal Observatory, Blackford Hill, Edinburgh EH9 3HJ, UK
        \and
            Department of Astronomy, Indiana University, Bloomington, Swain West, 727 E. 3rd Street, IN, 47405, USA}

  \abstract
    {
    In this paper, we present the results of a detailed photometric analysis of Glimpse-C02, one of the most extincted globular clusters of the Milky Way. We built a deep color magnitude diagram spanning $\approx 10$ magnitudes and enabling the very first identification of the main sequence turnoff of the cluster.
    Due to the extreme reddening conditions of the region where the stellar system is located, a differential reddening correction was necessary. 
    The resulting reddening map
    shows color excess variations up to $\delta E(B-V) \approx 2.5 $ mag in the direction of the target. 
    From isochrone-fitting of the differential reddening corrected color-magnitude diagram, we obtained a new estimate of the mean color excess, $E(B-V)=6.33^{+0.05}_{-0.04}$, and a distance modulus $(m-M)_0=14.00^{+0.26}_{-0.11}$, corresponding to a distance of $d=6.3^{+0.8}_{-0.3}$ kpc from the Sun, and a Galactocentric distance of $2.6^{+0.6}_{-0.7}$ kpc. This distance value, within the associated uncertainties, suggests that the cluster may be located closer to the Galactic Center compared to previous estimates, possibly supporting its classification as a bulge globular cluster. Furthermore, we obtained a photometric metallicity estimate of [Fe/H]$=-0.30^{+0.10}_{-0.08}$ and the first absolute age determination for Glimpse-C02, resulting in $t=11.9^{+0.7}_{-0.6}$ Gyr, as typically measured for Galactic globular clusters at this metallicity. 
    We also obtained a new estimate of the center of gravity of the cluster and determined its projected density profile from resolved star counts, finding  a high King concentration parameter ($c = 1.97_{-0.67}^{+0.51}$) and a core radius $r_c =8.72^{+0.40}_{-0.35}$ arcsec. Finally, from the surface brightness profile of the system, we derived an integrated $H$-band magnitude $M_{\rm H}=-7.9$, corresponding to a mass of $M=3.57^{+0.22}_{-0.19}\times 10^4 M_{\odot}$.
    Thus, our work classifies Glimpse-C02 
    as an old and metal-rich globular cluster that is in an advanced stage of its dynamical evolution.}

  \date{Received 03 March 2026 ; Accepted 29 March 2026}

   \maketitle

\section{Introduction}\label{Introduction}
The Milky Way (MW) hosts $\sim$170 globular clusters (GCs), a large fraction of which have been extensively studied, both photometrically and spectroscopically, using both space- and ground-based
facilities \citep[see, e.g.,][]{carretta+09, valenti10, kamann+18, libralato+22, ferraro+18b, ferraro+18a,ferraro+23,ferraro+26a}. As the oldest stellar systems in our Galaxy \citep[see, e.g,][]{marin+09, dotter+10, vandenberg+13, valcin+20, massari+23,aguado+25,ceccarelli+25}, they are key tracers of the early evolution of the MW, providing important insights into its formation and assembly history \citep{massari+19}. In this context, it is worth mentioning that two stellar systems, formerly classified as bulge GCs, Terzan 5 and Liller 1, have been found to exhibit properties that are incompatible with those of genuine GCs, such as the presence of multi-iron and multi-age
subpopulations \citep[][]{ferraro+09,ferraro+16,ferraro+21}. Their properties suggest that they could be fossil remnants of more massive structures that contributed to the formation of the bulge (see also \citealp{lanzoni+10, origlia+11}, \citeyear{origlia+13}, \citeyear{origlia+19},\citeyear{origlia+25}; \citealp[]{massari+14}; \citealp[]{pallanca+21b, dalessandro+22, crociati+23, crociati+24, deimer+24, fanelli+24, ferraro+25}) and could represent a significant source of gravitational waves \citep{ferraro+26b}. 
This reinforces the urgency to extend the investigation of 
Galactic GCs to include stellar systems that have been poorly studied so far because they are located in observationally challenging regions. In fact, a proper characterization of star clusters in regions such as the bulge and the disk has often been hampered by extreme obscuration due to interposed dust along the line of sight, and by severe stellar crowding. 

Near-infrared (NIR) surveys such as the Two Micron All Sky Survey \citep[2MASS;][]{skrutskie+06}, the VISTA Variables in the Via Lactea (VVV) survey \citep[][]{minniti+10} and its extension VVVX \citep[][]{minniti+16}, together with mid-infrared surveys, such as the Galactic Legacy Infrared Mid-Plane Survey Extraordinaire \citep[GLIMPSE;][]{benjamin+03} and the Wide-field Infrared Survey Explorer \citep[WISE;][]{wright+10}, have been fundamental for the study and the discovery of previously unknown GCs in the MW bulge and disk. Various methods have been employed to identify and confirm potential star clusters, for example, by building density maps to visually detect the presence of overdensities \citep[][]{minniti+17}. However, since not all over-densities correspond to real clusters, one of the most reliable approaches is represented by a kinematic analysis, especially exploiting the high precision of Gaia proper motions (\citealp[see, e.g.,][]{garro+20},\citeyear{garro+21}; \citealp[][]{obasi+21,minniti+21a}), to distinguish between cluster members and galactic field interlopers and to determine the real nature of the stellar system. 
Still, Gaia works mainly in the optical band, and the limiting magnitude in the G band is $\sim 20$, which is not enough to perform a complete kinematic analysis of the systems.
Moreover, only a fraction of the discovered stellar systems have been analyzed so far (\citealp[]{froebrich+07,borissova+14,minniti+17}, \citeyear{minniti+21a}, \citeyear{minniti+21b}; \citealp[]{camargo+18,palma+19,camargo+19,garro+20},\citeyear{garro+21}; \citealp[][among others]{obasi+21,dias+22,garro+22,kader+22,kader+23,kunder+24,hughes+26}) and current ground-based infrared (IR) facilities are not sufficient to completely characterize these objects. So, a very precise, high-resolution IR study is needed, exploiting, e.g., the superb capabilities of the Hubble Space Telescope (HST).

In this context, we present the first accurate photometric analysis of Glimpse-C02, one of the most extincted GCs of the MW. 
It was discovered in the GLIMPSE project \citep{benjamin+03}, which offered an excellent opportunity to explore a very extincted region of our galaxy, thanks to the use of NIR wavelengths, and to provide a deep census of reddened MW GCs. Glimpse-C02 is situated near the Galactic plane \citep[$l = 14^{\circ}.129$, $b = -0^{\circ}.644$; see N.3 in][]{mercer+05}, inside the thin disk, between its inner edge and the transition region with the bulge, at a distance $D\approx 4.6\pm 0.7$ kpc, based on the brightnesses of the red clump (RC) and the tip of the red giant branch (RGB) \citep[][]{kurtev+08}. It is a heavily extincted stellar system ($A_V\approx 24.8\pm 3$) that, according to moderate resolution K-band spectra of three probable members, has a metallicity [Fe/H]$=-0.33 \pm 0.14$ \citep[][]{kurtev+08}.
These results, however, have been obtained
with the ground-based instrument SofI at New Technology Telescope (NTT), which allowed the identification of just a portion of the RGB of the cluster, considering a region of $60''$ from the center \citep[see Figure 2 of][]{kurtev+08}. 
Hence, the high-resolution and NIR capabilities of the HST WFC3/IR camera represent an unprecedented opportunity to fully characterize this stellar system photometrically.

This work is organized as follows. Section \ref{Observations and Data Analysis} describes the NIR datasets used in the analysis and the adopted data analysis procedures. In Section \ref{NIR CMDs of Glimpse-C02}, we present the derived color-magnitude diagrams (CMDs) and discuss the main characteristics of the evolutionary sequences. Section \ref{Differential reddening} describes the method adopted to build the differential reddening map in the direction of the cluster. In Section \ref{Comparison with NGC 6440}, we show a comparison with a well-known bulge GC (NGC 6440) that is characterized by a similar value of metallicity, while Section \ref{Age and distance determination} is focused on the estimate of the absolute age of Glimpse-C02 through isochrone fitting and the determination of its distance modulus and color excess. In Section \ref{Determination of the structural parameters}, we present a new determination of the center of gravity and of the projected density and surface brightness profiles, which are used to derive the main structural parameters of the system via King model fitting, and to estimate the cluster total luminosity and mass. Finally, in Section \ref{Conclusions} we present a summary of the work and the main conclusions.

\section{Observations and data analysis}\label{Observations and Data Analysis}
   \subsection{NIR dataset}\label{NIR Data Set}
   
\begin{figure}
    \centering
    \includegraphics[width=9cm]{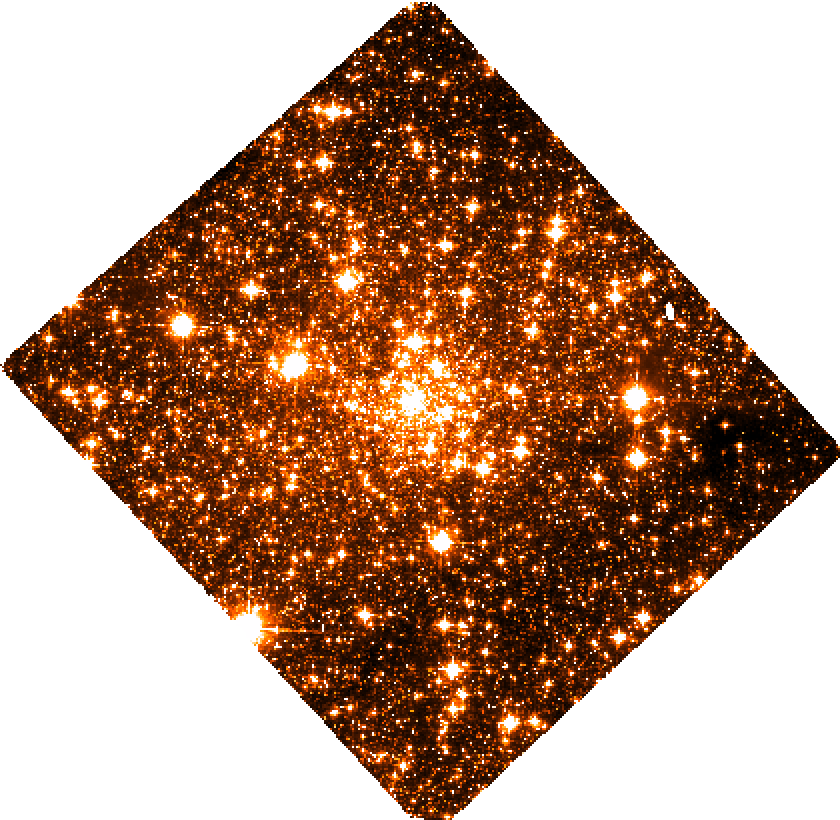}
    \caption{A WFC3/IR image centered on Glimpse-C02 obtained in the F160W filter, with an exposure time of 199 s. North is up, East is to the left.}
    \label{Glimpse_c02_f160}%
\end{figure}

The photometric investigation of Glimpse-C02 presented in this study is based on the analysis of the first HST epoch of this stellar system (proposal GO 17918, P.I. Loriga). It is composed of high-resolution images obtained from the IR channel of the WFC3/HST. This instrument provides $\sim$0.13 arcsec/pixel spatial resolution in a nominal $ 136 \times 123 $ arcsec$^2$ field of view (FoV). 
The data were acquired on the 30th of March 2025 and they are composed of 4 images obtained in F110W (2 x 399 s, 2 x 299 s) and 4 images in F160W with an exposure time of 199 s. 
For the sake of illustration, a F160W image of the studied cluster is shown in Fig.~\ref{Glimpse_c02_f160}.

To determine the stellar density profile along the entire radial extension of the system, we complemented the high-resolution HST data with a set of wide-field near-IR images obtained as part of the VVVX survey \citep{minniti+16}, as described also in \citet{Cadelano2022}. This dataset was acquired with the VISTA InfraRed CAMera (VIRCAM) mounted on the VISTA-ESO telescope. It is composed of 2 images obtained with the $J$ filter and an exposure time of 120 s, and 5 images with the $K_s$ filter and an exposure time of 16 s.

\begin{figure*}
    \centering
    \includegraphics[width=12cm]{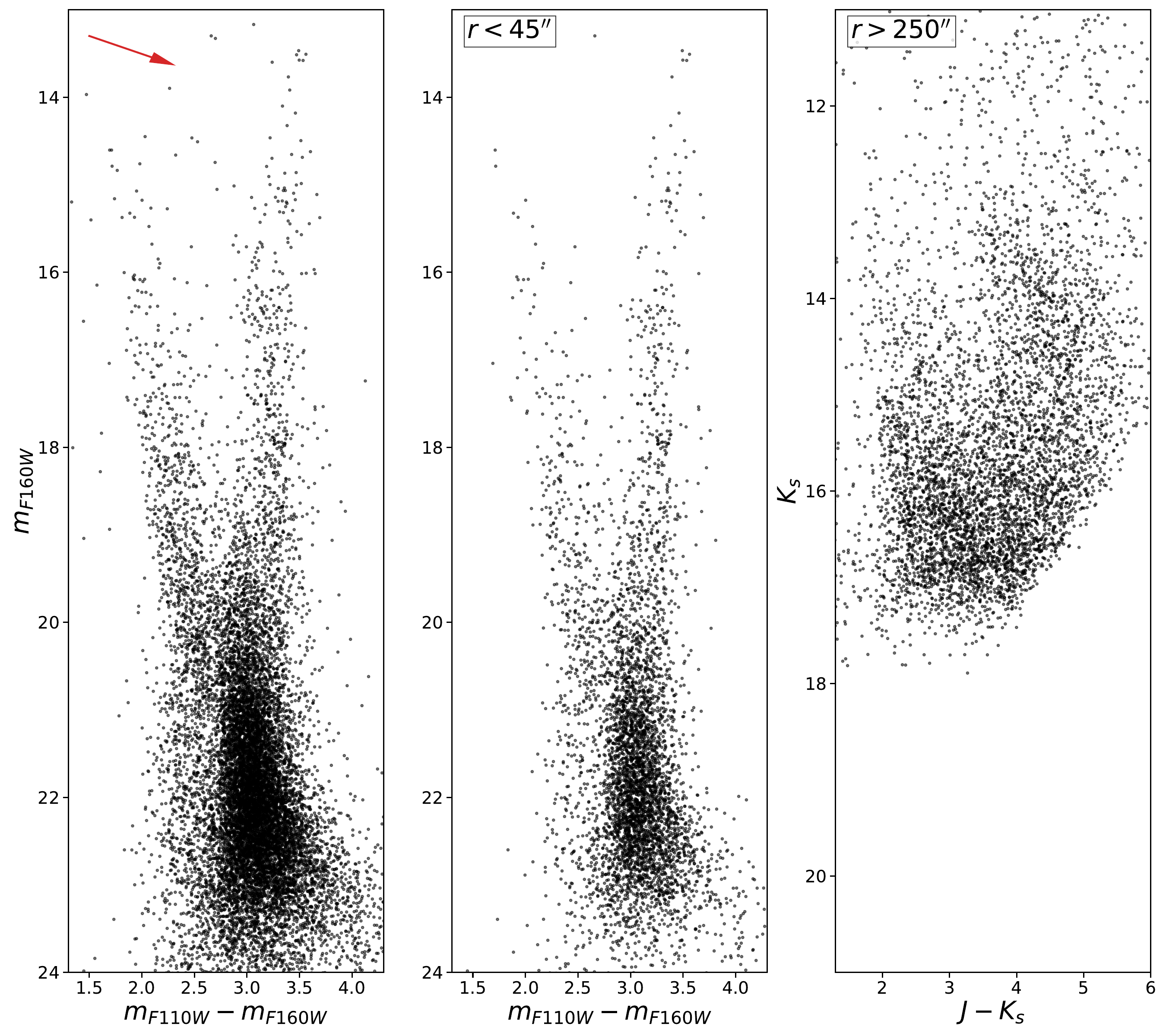}
    \caption{ 
    Left panel: ($m_{\rm F160W}$, $m_{\rm F110W}-m_{\rm F160W}$) observed CMD of Glimpse-C02 after the cleaning procedure using roundness, sharpness and chi parameters, with the reddening vector shown as a red arrow.
    Central panel: ($m_{\rm F160W}$, $m_{\rm F110W}-m_{\rm F160W}$) observed CMD of Glimpse-C02 after the same cleaning procedure and by considering a radius of $45 \arcsec$ from the center, as determined in Sec. \ref{Determination of the structural parameters}.
    Right panel: ($K_s$,$J-K_s$) CMD for distances beyond $250 \arcsec$ from the center, obtained from the analyzed VIRCAM data.}
    \label{cmds_ir}
\end{figure*} 

    \subsection{Data reduction}\label{Data reduction}
In our photometric analysis, we used {\it flt} images, which are already processed and calibrated (i.e., flat-fielded, dark-subtracted, etc.) by the Space Telescope Science Institute (STScI) pipeline, as well as {\it drz} images, which are calibrated and corrected for geometric distortions.
The analysis of the dataset has been performed using DOLPHOT \citep[][]{dolphin+2000,dolphin+16}, a stellar photometry package that was originally created to obtain accurate stellar photometry PSF-fitting with the Wide Field and Planetary Camera 2 (WFPC2) data, but it was then generalized to use analytic PSF models for any camera. The procedure is very similar to the classic PSF-fitting routine and it includes also an initial image preparation, meaning a masking of bad columns, cosmic-ray cleaning, hot pixel masking, etc.; a sky or background determination and a final aperture corrections evaluation and application to provide a calibrated photometry. DOLPHOT also performs Pixel Area Map (PAM) and Charge Transfer Efficiency (CTE) corrections. In our procedure, we used the {\it drz} F160W image as the reference frame for all the other F160W and F110W images, for which the geometric transformations were determined. We set a threshold of $5\sigma$ 
for inclusion of the stars in the final catalog. 
All the remaining parameters were set to the default values for the camera. The final product is a catalog in which magnitudes from different images are combined using a weighted mean that accounts for the $\chi^2$ of the PSF-fitting procedure. Moreover, the magnitudes are reported to the VEGAMAG photometric system, using the appropriate zero points and encircled energy fractions, and including aperture corrections. Furthermore, the final catalog contains several photometric quality flags (e.g., chi, sharpness, roundness) that are useful for preliminary cleaning to remove poorly measured stars.
The final necessary step to build the final catalog was to transform the instrumental coordinates to the absolute reference frame ($\alpha$ and $\delta$). To do that, we used the stars in common with the publicly available 2MASS catalog \citep[][]{skrutskie+06}, which was used as a sort of 'bridge catalog' to obtain the astrometric information from the Gaia DR3 catalog \citep{gaia+23}, through the cross-correlation software \texttt{CataXcorr} \citep{montegriffo+95}. 
In the case of the VIRCAM data, the data reduction was performed using a standard approach
suitable for ground-based observations with the DAOPHOT
and ALLFRAME packages \citep[see, e.g.,][]{Loriga2025}.
The resulting catalog was astrometrized using the stars in common with the Gaia DR3 catalog \citep{gaia+23} and the instrumental
magnitudes were calibrated using the stars in common with the 2MASS catalog \citep[][]{skrutskie+06}, obtained with the same filters and in an overlapping region of the sky. The derived CMD is shown in the right panel of Fig.~\ref{cmds_ir}.

\subsection{NIR CMD of Glimpse-C02}
\label{NIR CMDs of Glimpse-C02}    
The CMD obtained from the analyzed observations is shown in the left panel of Fig.~\ref{cmds_ir}, after applying a cleaning procedure based on the chi, sharpness, and roundness parameters, using a $3\sigma$ clipping criterion.
The central panel of the same figure shows the CMD including only stars located within a radius of $45\arcsec$ from the system’s center of gravity, as defined in Sec.~\ref{Determination of the structural parameters}.
This represents the first CMD of Glimpse-C02 ever obtained from high-resolution HST data, and its comparison with previous results published in the literature \citep[see Fig.2 of][]{kurtev+08} allows us to fully appreciate the advantages of the use of space-based high-resolution IR data.
By analyzing the CMD, it is possible to identify, for the first time for this stellar system, the main sequence turn off (MS-TO), which is located around $m_{\rm F160W} \approx 20$ ($m_{\rm F110W} \approx 23$).
The CMD extends $\approx$3 mag below it. However, the effect of differential reddening is evident as a huge spread of all the evolutionary sequences, 
which are also affected by intense field contamination, as clearly visible mainly in the left part of the CMD. Unfortunately, it is not possible to decontaminate the cluster population from field interlopers.
In fact, no second epoch observations exist to measure relative proper motions and discriminate cluster members from Galactic field interlopers. We also searched for stars in common with Gaia DR3 \citep{gaia+23}, but only a few matches with measured proper motions were found, mostly located in CMD regions typical of field stars, thus preventing a reliable decontamination. In addition, no complementary observations of the field surrounding the cluster are available to perform a statistical decontamination \citep[e.g.,][]{dalessandro08,Dalessandro2019,Giusti2023_kmk,rosignoli+25}.

Despite all this, the presence of a well-populated and extended RGB, which is a characteristic commonly associated with old stellar systems, can be clearly recognized in the CMD. Moreover, the lack of a blue horizontal branch suggests a likely metal-rich GC, in agreement with the spectroscopic analysis of \citet[][]{kurtev+08}. A low-significance over-density along the RGB at magnitudes $m_{\rm F110W}\approx19.5$ and $m_{\rm F160W}\approx16.5$ suggests the presence of the red clump typical of metal rich-populations.

\section{Differential reddening}\label{Differential reddening}
One of the main challenges in characterizing Glimpse-C02 is represented not only by the presence of strong interstellar extinction, but also by its spatial variability along the line of sight. This is known as differential reddening and it is due to the presence of clouds with different column densities on scales as small as a few arcsec \citep[see, e.g.,][]{pallanca+21b}. 
As a consequence, the amount of color excess $E(B-V)$, defined as the difference between the observed color $(B-V)$ and the intrinsic one $(B-V)_0$, varies from star to star depending on their position within the observed field. The primary effect on the CMD is a stretching of the evolutionary sequences along the reddening vector. To correct the CMDs for this effect, the first step consists of the assumption of the "reddening law" \citep[see, e.g.,][]{cardelli+89,fitzpatrick+90,odonnell+94,fitzpatrick+99}, meaning the wavelength dependence of interstellar extinction relative to the absolute extinction in the $V$-band.
Indeed, the direction of the reddening vector in a CMD depends on the adopted photometric filters, through wavelength-dependent parameters commonly referred to as $R_\lambda$. These parameters are related to the extinction coefficients via the relation $A_{\lambda} = R_{\lambda} \times E(B-V)$. Each $R_\lambda$ can be expressed as the product of two terms: $R_\lambda = R_V \times c_{\lambda, R_V}$.
Here, $c_{\lambda, R_V}$ corresponds to the ratio $A_\lambda / A_V$ (or equivalently, $R_\lambda / R_V$), which describes the reddening law. In our case, we assume the standard "reddening law" described by \citet{cardelli+89} with $R_V=3.1$, corresponding to $R_{\rm F110W}=1.04345$ and $R_{\rm F160W}=0.63296$.

\begin{figure}
   \centering
    \includegraphics[scale=0.25]{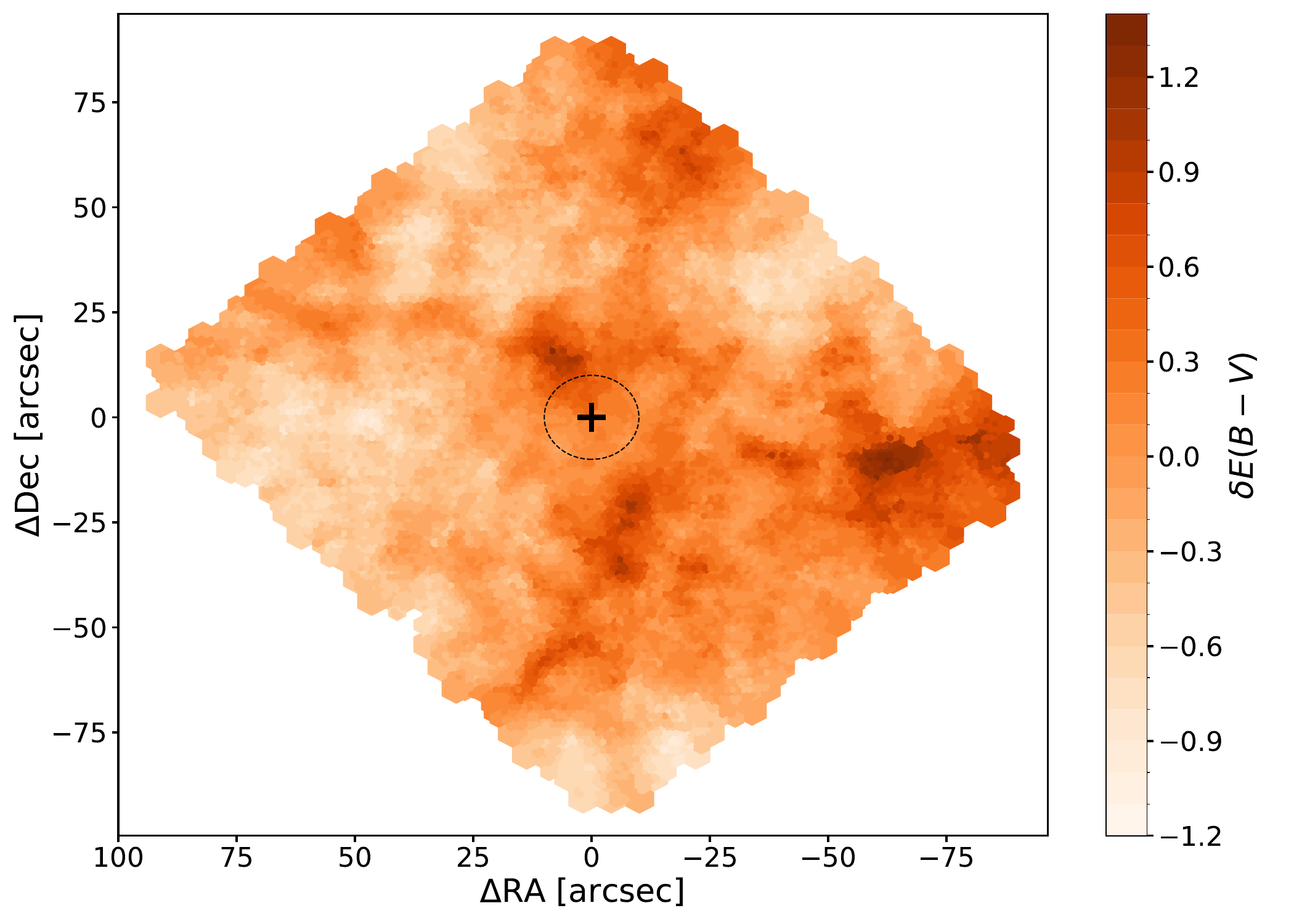}
   \caption{Differential reddening map relative to the cluster center position (black cross; see Sec.\ref{Determination of the structural parameters}). 
   North is up, East is to the left. Darker colors correspond to more extincted regions, as detailed in the side color bar that reports the differential color excess $\delta(B-V)$.}
    \label{reddening_map}
\end{figure}

To perform the differential reddening correction, we adopted a star-by-star approach, described in detail by \citet[see also \citealp{cadelano+20,pallanca+21b,Loriga2025}]{pallanca+19}. The method consists in determining the mean ridge line (MRL) of Glimpse-C02 in the NIR CMD by selecting a sample of well-measured, likely member stars located along the RGB, SGB, and upper MS. For each star in our HST catalog, we then identified a group of nearby $N_*=30$ stars (selected within a maximum radius of $10\arcsec$) that were also well measured. This group, composed of $N_*$ stars, was used to construct a so-called “local CMD”.
To estimate the differential reddening affecting each star, we shifted the MRL along the reddening vector in steps of $\delta E(B-V)$ until it matched the local CMD. 
The differential reddening map obtained from this procedure is shown in Fig.~\ref{reddening_map} on a FoV of roughly $ 136 \times 123 $ arcsec$^2$, corresponding to the nominal FoV of the WFC3. In the figure, only the regions with reliable differential reddening corrections are shown.
As can be seen, the reddening appears patchy, with different regions characterized by different reddening features. On the western side of the FoV it is possible to distinguish a highly reddened region (darker colors).
The same holds for the southern side, where a more elongated filament is observed, and in the north-west part of the map. 
A small dark blob can also be noticed on the north with respect to the center of the cluster (black cross; see Section \ref{Determination of the structural parameters}).
Some lighter blobs, meaning smaller values of differential reddening, can be seen in the north-west region of the FoV, as well as in the east part.
Overall, the map graphically indicates an extremely large variation in the extinction, reaching a maximum value of $\delta E(B-V) \approx 2.5$ mag in the whole FoV. 

\begin{figure*}[hbt!]
   \centering
 \includegraphics[width=14cm]{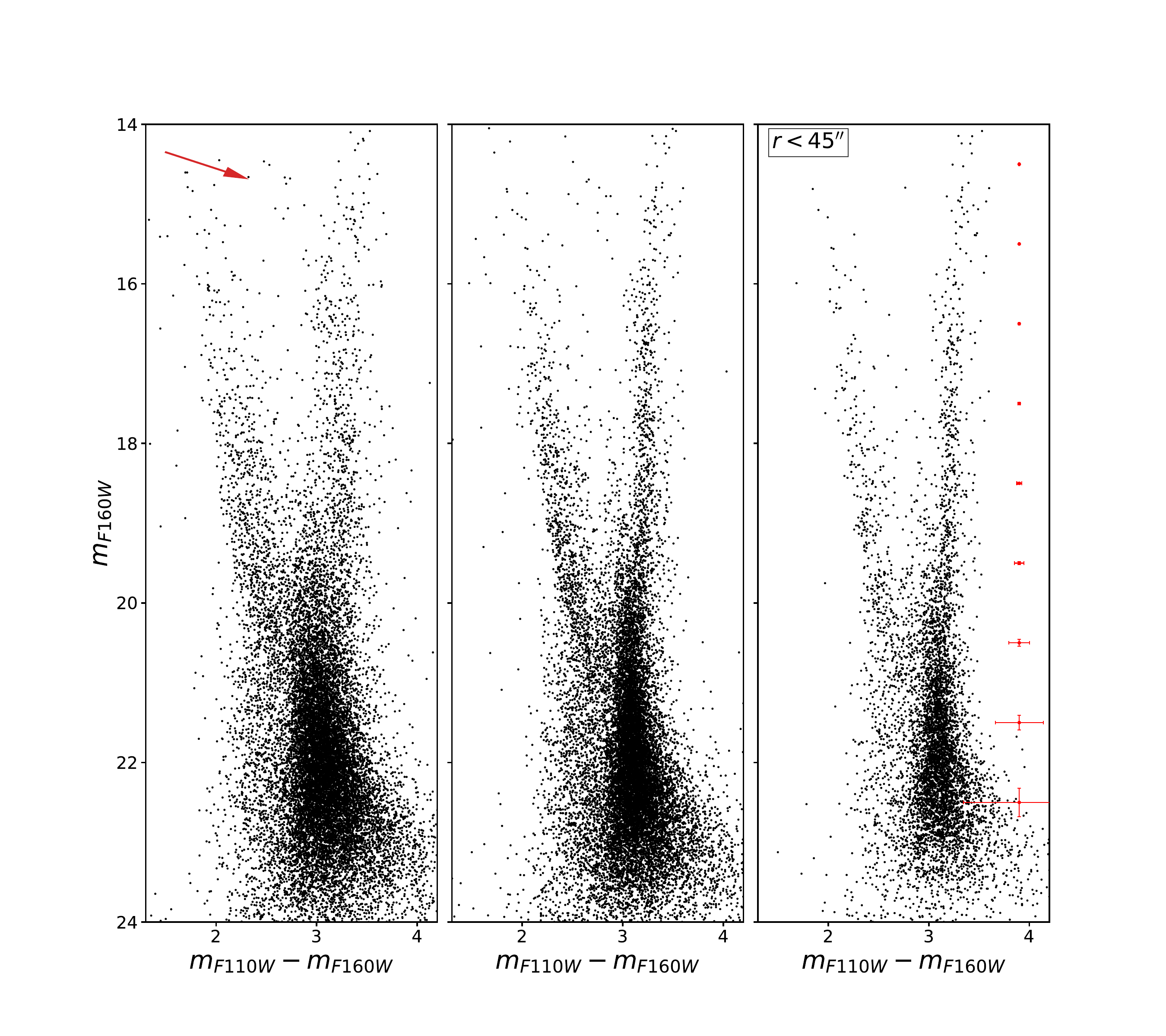}
   \caption{
   ($m_{\rm F160W}$, $m_{\rm F110W}-m_{\rm F160W}$) 
   CMD before the correction for differential reddening (left panel), after the correction for differential reddening (central panel), and after the correction for differential reddening by considering only the stars within $45\arcsec$ from the center, as defined in Sec.\ref{Determination of the structural parameters} (right panel). The reddening vector is represented in the left panel as a red arrow. The red crosses in the right panel represent the $3\sigma$ mean uncertainties on the color and the magnitude for each magnitude bin.}
    \label{tripanel_redd_corr}
\end{figure*}

The derived reddening map has been applied to correct the effects of differential reddening in the observed CMDs (see Fig.~\ref{tripanel_redd_corr}). A residual differential reddening is likely still present in the CMD, as shown in the right panel of Fig.~\ref{tripanel_redd_corr}, where the red crosses indicate the $3\sigma$ mean magnitude and color uncertainties in each magnitude bin, which are smaller than the residual width of the sequences. 
Overall, however, the significant improvement achieved through the correction is clearly visible, particularly in the RGB of the central panel of Fig.~\ref{tripanel_redd_corr}, as well as in the MS-TO region, which appears noticeably sharper and better defined.

\section{Comparison with NGC 6440}\label{Comparison with NGC 6440} 
\begin{figure}
    \centering
    \includegraphics[scale=0.20]{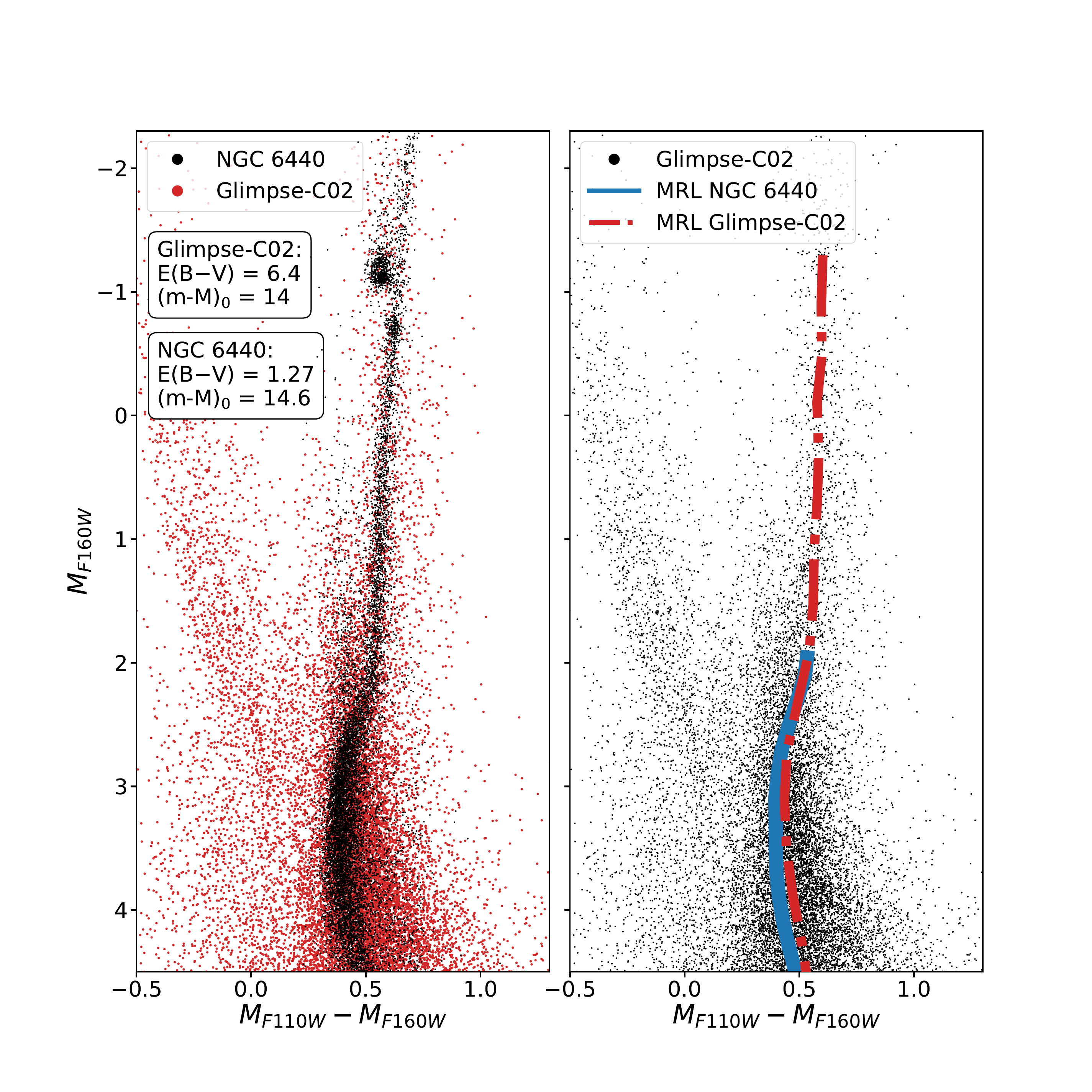}
    \caption{Left panel: superposition of the $(M_{\rm F160W}$, $M_{\rm F110W}-M_{\rm F160W}$) CMD of Glimpse-C02 (red dots) with the $(M_{\rm F160W}$, $M_{\rm F110W}-M_{\rm F160W}$) CMD of NGC 6440 (black dots) in the absolute plane. The parameters used for each GC are labeled. Right panel: $(M_{\rm F160W}$, $M_{\rm F110W}-M_{\rm F160W}$) CMD of Glimpse-C02 in the absolute plane with the MRL of NGC 6440 (blue line) and Glimpse C02 (red line) superposed.}
    \label{glimpse_meanridgeline}
\end{figure}

    First of all, given the complexity of Glimpse-C02, we adopt a qualitative approach that consists in comparing the CMD of this system with that of the well-studied bulge GC NGC 6440. The comparison can provide first-guess values of the distance and mean color excess of Glimpse-C02. Indeed, NGC 6440 provides a useful reference, as its metallicity \citep[$\mathrm{[Fe/H]}=-0.56$,][]{origlia+97,origlia+08} is similar to that of Glimpse-C02. Moreover, NGC 6440 was observed with the same HST WFC3/IR filters as Glimpse-C02, thus allowing a straightforward comparison between the two. In particular, the dataset (Proposal GO 12517, P.I. F.R. Ferraro) is composed of 8 images obtained in the F160W filter (1 x 26 s and 7x 249 s) and 8 images in 
    F110W (1 x 26 s and 7x 349 s). For this cluster, we performed the same reduction procedure of Glimpse-C02 described in Sec. \ref{Data reduction}, by using an F160W drz image as a reference. 
    The distance modulus of NGC 6440 ($ (m-M)_0 = 14.6 $), together with its age ($t = 13$ Gyr) and reddening ($E(B-V) = 1.27$), have been accurately determined by \citet[][see also \citealt{Cadelano2017,Cadelano2023}]{pallanca+21a}. Therefore, by using the differential reddening-corrected CMD of Glimpse-C02 in the absolute plane, using the values of distance modulus and color excess from \citet{kurtev+08}, we evaluated the MRL of the system, considering only stars located around the HB, RGB and the upper part of the MS, meaning most likely belonging to the cluster. We then constructed the MRL of NGC 6440 and we shifted the one of Glimpse-C02 until they matched (right panel of Fig.\ref{glimpse_meanridgeline}).The values that minimize the color and magnitude difference between the two MRLs have been used to obtain first-guess estimates of the distance modulus and mean color excess for Glimpse-C02: $(m-M)_0=14.0$, $E(B-V)=6.4$
    (left panel of Fig.\ref{glimpse_meanridgeline}).
   The very good match between the two CMDs and MRLs suggests that Glimpse-C02 has an age and a metallicity comparable to those of NGC 6440. 

\section{Age and distance determination}\label{Age and distance determination}

\begin{figure*}
\centering

\begin{minipage}{0.48\textwidth}
    \centering
    \includegraphics[scale=0.37]{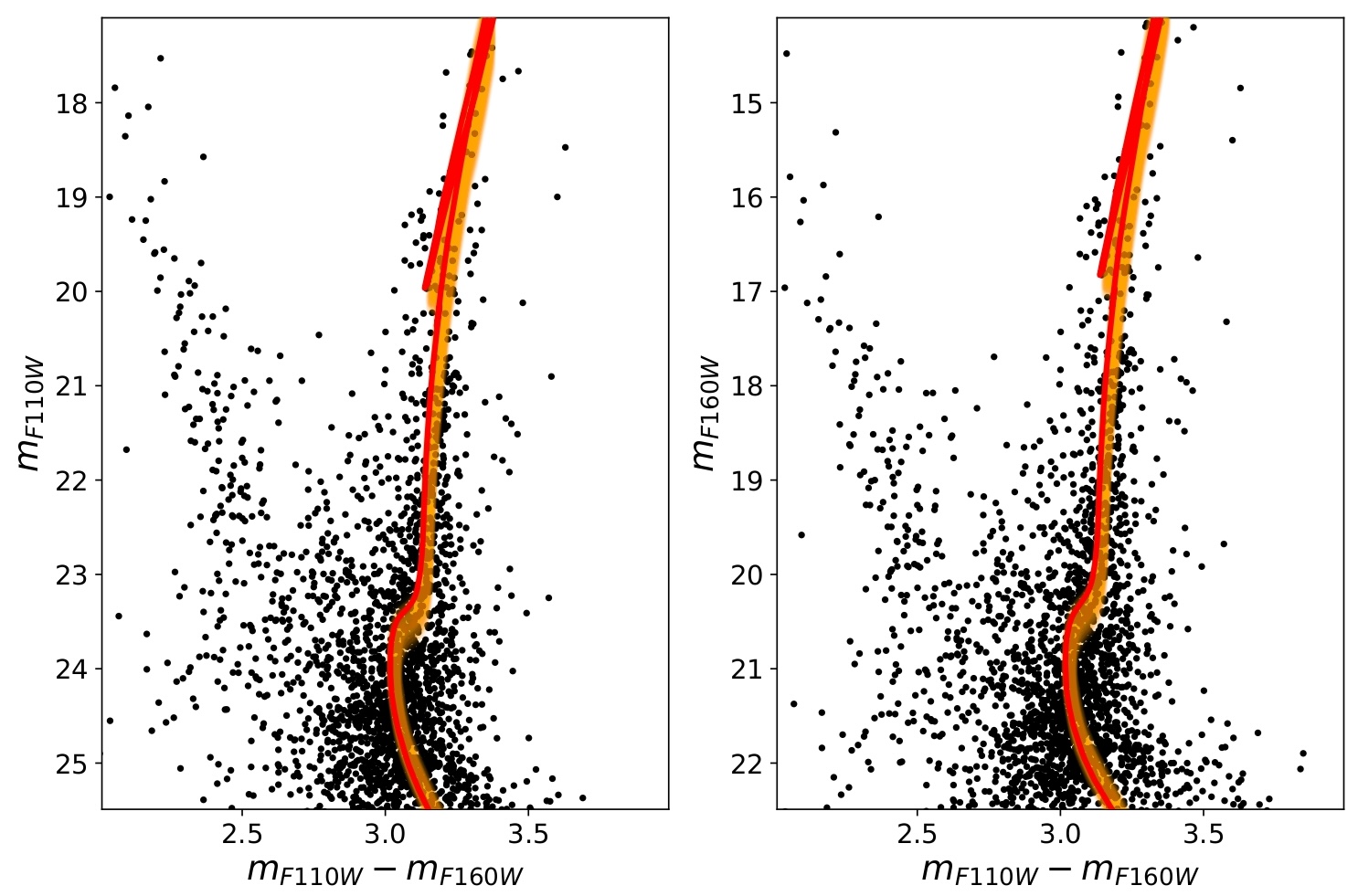}
\end{minipage}
\hfill
\begin{minipage}{0.48\textwidth}
    \centering
    \includegraphics[width=\linewidth]{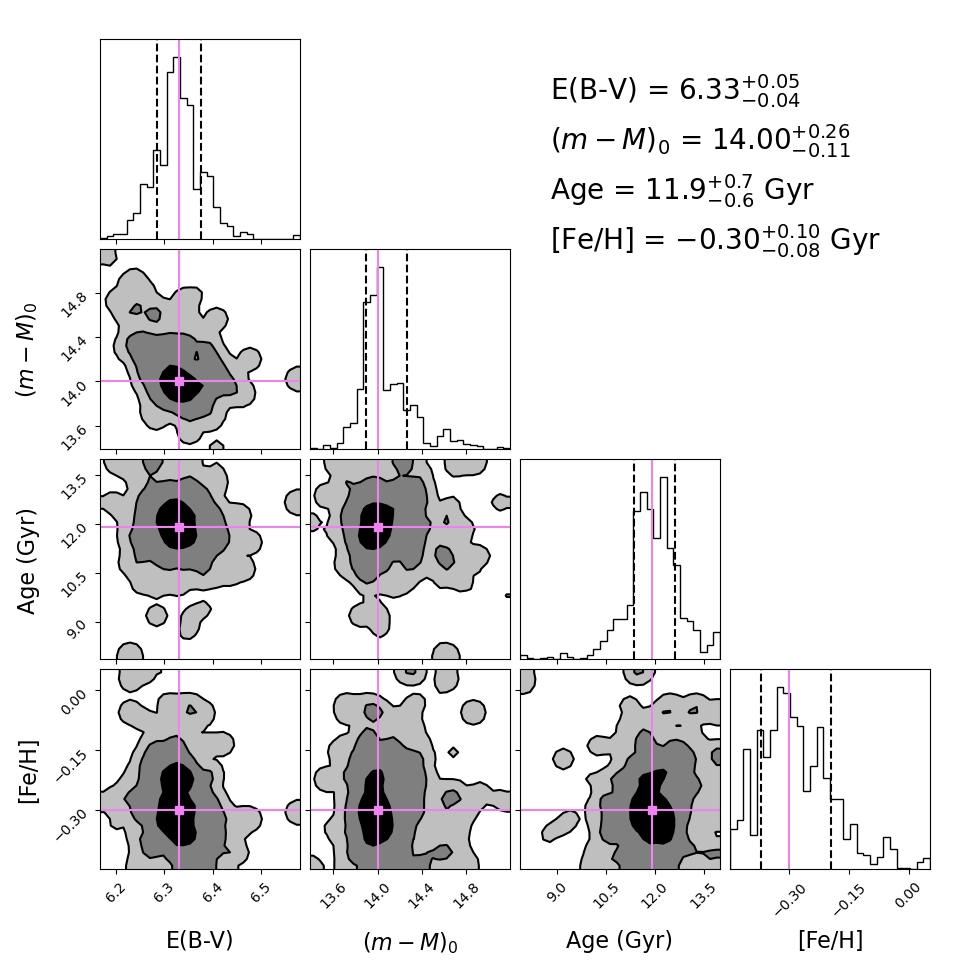}
\end{minipage}

\caption{Left: Differential reddening corrected  ($m_{\rm F110W}$, $m_{\rm F110W}-m_{\rm F160W}$), ($m_{\rm F160W}$, $m_{\rm F110W}-m_{\rm F160W}$) CMDs of stars within $25\arcsec$ from the cluster center and with over-plotted the BASTI isochrone (red line) that best reproduces the evolutionary sequences of Glimpse-C02. The orange shaded region represents the 1$\sigma$ uncertainty region of the best-fit isochrone. Right: Corner plot with the one- and two-dimensional projections of the posterior probability distributions for all the parameters
as obtained from the BaSTI isochrones. The contours correspond to the 1$\sigma$, 2$\sigma$,
and 3$\sigma$ levels. The values of the color excess, distance modulus, age and metallicity providing the best match with the data are also labeled. }
\label{age}
\end{figure*}

    Starting from the first-guess values of the distance modulus and reddening of Glimpse-C02 obtained from the superposition of its CMD onto that of NGC 6440, we refined both the estimates and determined the age and metallicity of the cluster via the isochrone fitting technique. In this approach, the observed CMD of the cluster is compared with a set of theoretical isochrones to simultaneously estimate age, reddening, distance modulus, and metallicity. To explore the parameter space and derive the best-fit solution, we used the Monte Carlo Markov Chain (MCMC) approach described in \citet[see also \citealp{deras+23,deras+24,Giusti2024,giusti+25}]{cadelano+20}, implemented through the emcee code \citep{foremanmackey+13, foremanmackey+19}.
    We adopted the BASTI isochrones \citep{hidalgo+18,pietrinferni+21,pietrinferni+24}, with [$\alpha$/Fe]$=0.4$, a standard He abundance of $Y=0.247$, and including overshooting as well as RGB mass loss. We assumed flat priors for the age, spanning 
    from 2 to 14 Gyr in steps of 0.1 Gyr, and metallicities in the range $-1<[{\rm Fe/H}]<0.2$ with steps of 0.05  dex.  
    Moreover, Gaussian priors centered on the values obtained from the comparison with NGC 6440, with a $\sigma=0.1$, were adopted for $E(B-V)$ and $(m-M)_0$.   
    The left panel of Fig. \ref{age} shows the isochrone that best reproduces the evolutionary sequences
    for $r<25\arcsec$ from the center of the system.
    The complete set of parameters derived by the procedure are: $(m-M)_0=14.00^{+0.26}_{-0.11}$, $E(B-V)=6.33^{+0.05}_{-0.04}$, $t=11.9^{+0.7}_{-0.6}$ Gyr, and [Fe/H]$=-0.30^{+0.10}_{-0.08}$.
    The results robustly confirm that Glimpse-C02 is an old system, fully consistent with the indication already suggested by the visual inspection of its CMD.
    Our estimate of the distance ($d=6.3^{+0.8}_{-0.3}$ kpc) appears to be slightly larger than the value of $d=4.6\pm 0.7$ kpc published in \citet{kurtev+08}. The corresponding distance from the Galactic center is $2.6^{+0.6}_{-0.7}$ kpc, smaller than the literature one.
    The mean color excess is significantly smaller than the literature value, namely, $E(B-V)=7.85$ \citep{harris+96,kurtev+08}. A good agreement  within the uncertainties is found, instead, for the derived value of [Fe/H].
   
\section{Determination of the structural parameters}\label{Determination of the structural parameters}
\subsection{Radial density profile}\label{Radial density profile}
    The first step in determining the cluster's density profile is to estimate the system's center of gravity ($C_{\rm grav}$). In this study, we derived it using resolved star counts, made possible by the high spatial resolution of the data.
    To determine $C_{\rm grav}$, we computed the barycenter of stars selected within specific magnitude range and radial distances, starting from an initial guess value \citep{harris+96}. This procedure was repeated iteratively, using the barycenter obtained from the previous iteration as the new center. Convergence was achieved when the difference between two successive estimates was less than $0.01\arcsec$ \citep[see, e.g.,][]{montegriffo+95, miocchi+13,Giusti2024}.
    The analysis was carried out by considering stars selected within 4 different radial ranges (30$\arcsec$, 35$\arcsec$, 40$\arcsec$, 45$\arcsec$) and 3 magnitude ranges ($20.75<m_{\rm F160W}<16.0$, $20.5<m_{\rm F160W}<16$, $19.5<m_{\rm F160W}<16.0$), restricting the sample to the color range $2.8<(m_{F110W}-m_{F160W})<3.5$, in order to mitigate field contamination. Radial distances were chosen to sample the gradient in the surface brightness of the cluster, which becomes noticeable at approximately $\approx 10\arcsec$ based on a visual inspection of the maps. The faint magnitude limits were selected to guarantee high statistics while avoiding spurious fluctuations due to photometric incompleteness.
    The bright magnitude cut at $m_{\rm F160W} = 16$ was applied to exclude saturated stars. The final position of $C_{\rm grav}$ was obtained by averaging the 12 barycenter estimates, while its uncertainty was estimated as the standard deviation of the mean. The coordinates of $C_{\rm grav}$ thus determined are: $\alpha_{\rm ICRS} = 18^{\rm h} 18^{\rm m} 30.54^{\rm s}$, $\delta_{\rm ICRS} = -16^{\circ} 58\arcmin 34.94\arcsec$, with uncertainties of $0.09\arcsec$ in right ascension and $0.25\arcsec$ in declination.
    This new estimate differs by $3.12\arcsec$ from the value quoted by \citet{harris+96}.
    
\begin{figure}
    \centering
    \includegraphics[width=8cm]{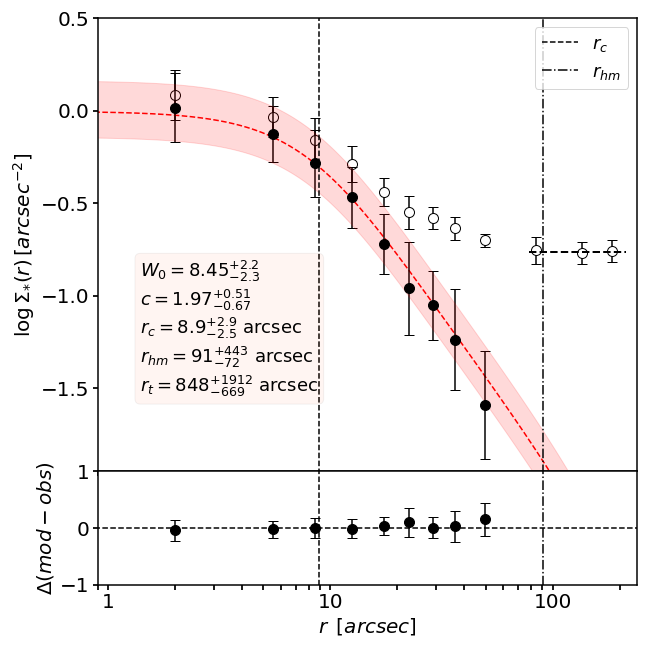}
    \caption{Projected density profile of Glimpse-C02 derived from star counts in concentric annuli around $C_{\rm grav}$ (empty circles) by combining WFC3 and VIRCAM data. The horizontal dashed segment indicates the Galactic field density, which has been subtracted from the observed points (empty circles) to obtain the background-subtracted profile, shown as filled circles. The red solid line represents the best-fit King model to the cluster's density profile, with the red stripe indicating the $\pm 1 \sigma$ range of solutions. The vertical lines denote the positions of the core radius (short dashed line), half-mass radius (dot-dashed line). The values of the main structural parameters obtained from the fitting process are 
    labeled (see details in the text).} 
    \label{dens_prof}
\end{figure}

    After the determination of $C_{\rm grav}$, we proceeded to derive the radial density profile based on resolved star counts \citep[see, e.g.,][]{miocchi+13,Cadelano2017_m71,Beccari2023,Loriga2025}. To sample the inner $\approx 60\arcsec$ from the center, we used the WFC3 dataset. For the cluster’s outer regions ($60\arcsec \leq r \leq 210\arcsec $), we used the wide-field VIRCAM dataset.
    In both cases, the FoV was divided into concentric rings of different sizes, chosen to ensure good statistics. Within each ring, we calculated the stellar density as the number of stars divided by the area. To minimize incompleteness effects in the inner regions, we considered only stars with $m_{\rm F160W} < 20$, with a cut performed along the direction of the reddening vector.
    For the VIRCAM wide-field data, we considered only stars with a magnitude $K_{s} \leq 14$.  The VIRCAM profile was then renormalized, by anchoring it to the two outermost radial bins of the WFC3 profile, which are in common between the two profiles.
    The projected density profile obtained from this analysis is displayed in Fig. \ref{dens_prof} as empty circles. In the outermost regions, at only $60\arcsec$ from the cluster center, the profile exhibits a flattening, with the logarithm of the density remaining roughly constant at $\sim -0.7$ stars ${\rm arcsec}^{-2}$. This is the signature of the intense contamination by Galactic field stars. 
    This value was then subtracted from the observed profile, and the decontaminated radial density distribution of the cluster is shown as filled circles in Fig.\ref{dens_prof}. 
    To derive the structural parameters of Glimpse-C02, we adopted the fitting procedure described in \citet{raso+20,Cadelano2022}. 
    Specifically, we performed an MCMC fitting of the observed density profile using \citet{king+66} models, with flat priors on the fitting parameters, namely the central density, the concentration parameter ($c$), and $r_c$. Moreover, a $\chi^2$ likelihood function was used to derive the best fitting parameters.
    The resulting best-fit model is shown in Fig.~\ref{dens_prof} and is characterized by a concentration parameter $c = 1.97_{-0.67}^{+0.51}$, corresponding to a dimensionless central potential $W_0 = 8.45_{-2.3}^{+2.2}$, and a core radius $r_c = 8.9_{-2.5}^{+2.9}$ arcsec,
    a half-mass radius $r_{hm} = 91_{-72}^{+443}$ arcsec, and a tidal radius 
    $r_t = 848_{-669}^{+1912}$ arcsec.

\subsection{Surface brightness profile}\label{surface brightness profile}
\begin{figure}
    \centering
    \includegraphics[width=8cm]{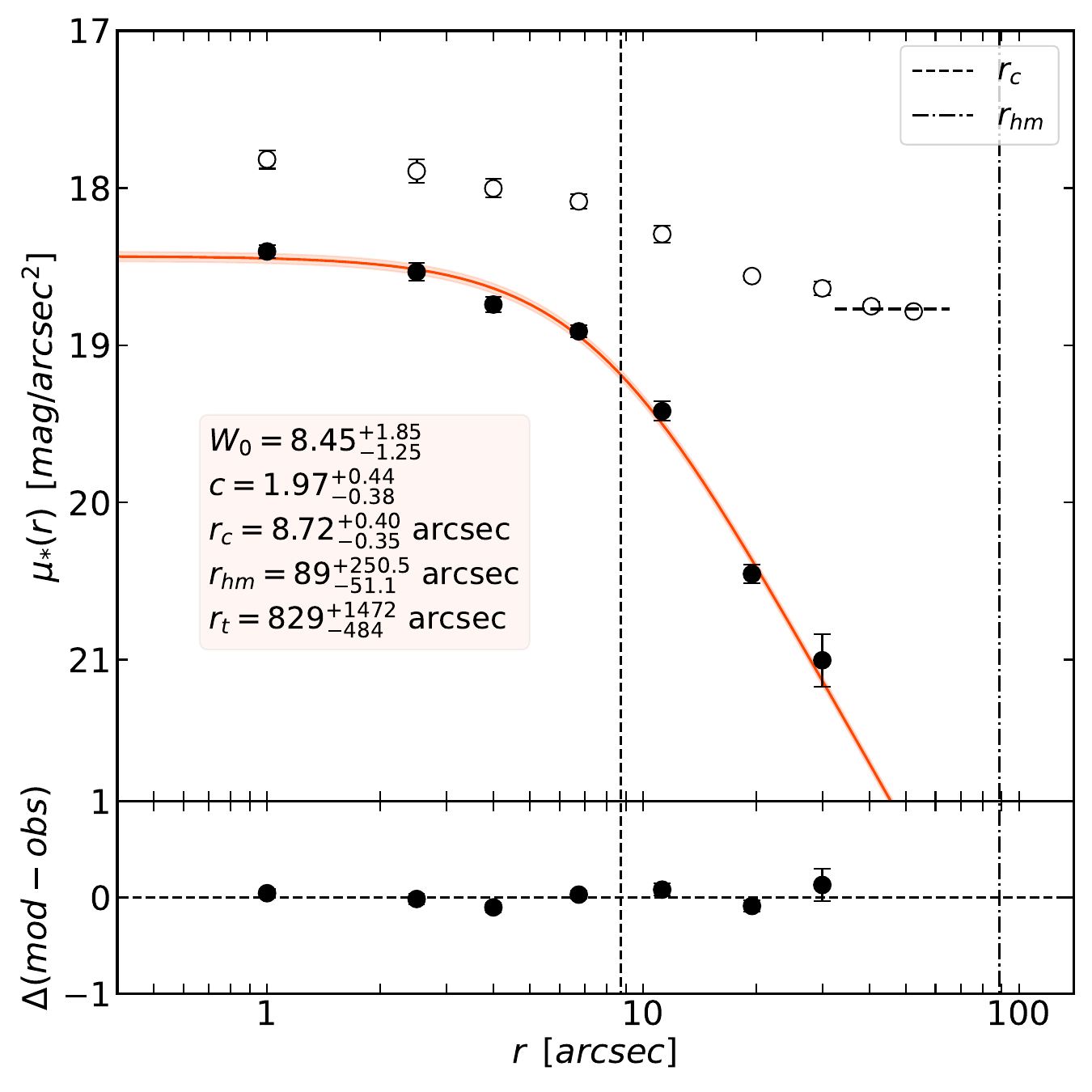}
    \caption{Same as in Fig. \ref{dens_prof} but for the surface brightness profile in the HST/WFC3 F160W filter.} 
    \label{sb_prof}
\end{figure}

To double-check the values of the structural parameters and estimate the cluster total luminosity and mass, we also analyzed its surface brightness profile. The procedure closely follows that adopted for the stellar density profile (see Sec. \ref{Radial density profile}). The WFC3 FoV of a \textit{drz} F160W-band image was divided into a set of concentric annuli, and each annulus was further subdivided into four angular sectors. The integrated surface brightness of each ring was computed as the average of the four sector values, while the associated uncertainty was estimated as their standard deviation. The resulting surface brightness profile is shown in Fig. \ref{sb_prof} (empty circles). The mean surface brightness of the background was estimated from the average of the outermost two data points, yielding a value of $\mu_{\rm {F160W}}\approx18.8$ mag arcsec$^{-2}$. This background level was subtracted from all measurements, producing the decontaminated surface brightness profile (filled circles). The corrected profile was fitted following the same methodology described above, through comparison with a grid of King models. The best-fitting parameters are a dimensionless central potential $W_0 = 8.45_{+1.85}^{-1.25}$, a King concentration index $c = 1.97^{+0.44}_{-0.38} $, a central surface brightness $\mu_{F160W,0} = 18.43^{+0.03}_{-0.03}$ mag arcsec$^{-2}$, a core radius $r_c =8.72^{+0.40}_{-0.35}$ arcsec , a half-mass radius $r_{hm} = 89^{+250.5}_{-51.1}$ arcsec, and a tidal radius $r_t = 829^{+1472}_{-484}$ arcsec. All these parameters are consistent, within the uncertainties, with those derived from the fit to the stellar density profile. It is worth emphasizing that the parameters $W_0$ and $c$ match exactly.

To estimate the integrated absolute magnitude of the system, we assumed the distance modulus and color excess previously determined (Sec. \ref{Age and distance determination}), $R_{\rm F160W} = 0.63296$ (see Sec. \ref{Differential reddening}), an $H$-band mass-to-light ratio for a Kroupa initial mass function of $M/L_{\rm H}=1.2$ \citep{maraston+00,maraston+05}, and an absolute magnitude in the F160W filter for the Sun of $M_{\odot,\rm F160W}=3.36$ \citep{willmer+18}. The integration of the surface brightness profile under these assumptions provided us with an integrated absolute magnitude in the $H$ band of $M_{\rm H}=-7.9$, which corresponds to a luminosity $L_{\rm H}=3.05^{+0.19}_{-0.16}\times10^4 L_{\odot}$, and a total mass of $M=3.57^{+0.22}_{-0.19}\times 10^4 M_{\odot}$ for Glimpse-C02.
We also computed the cluster's half-mass relaxation time ($t_{\rm rh}$) following \citet{Spitzer+71}. Using the cluster mass in solar masses, the distance derived in Sec.~\ref{Age and distance determination}, 
and the half-mass radius from the fit to the density profile (Sec.~\ref{Radial density profile}), 
we obtained $\log(t_{\rm rh}/{\rm yr}) = 8.7$. Then, by considering the estimated age (Sec. \ref{Age and distance determination}), we evaluated $t/t_{rh}=23.6$.

\section{Conclusions}\label{Conclusions}
This work presents the first comprehensive photometric study of Glimpse-C02, one of the most heavily obscured GCs in the Galaxy. By taking advantage of the first available HST/WFC3 IR 
data, we conducted a detailed analysis of its stellar population. 
We constructed a deep NIR CMD extending by $\approx 10$ mag and reaching $\approx 3$ mag below the MS-TO. This represents the first CMD that clearly reveals the position of the MS-TO of the cluster, providing an unprecedented view of its stellar population.
Because the system lies in a highly reddened region, we derived a high-resolution reddening map that allowed us to perform a differential reddening correction, revealing variations up to $\delta E(B-V) \approx 2.5$ mag across the entire FoV. 
Despite the use of IR filters and well-tested differential reddening correction techniques, proven to be effective even in the most challenging cases, such as NGC 6440 \citep[]{pallanca+21a} and Liller 1 \citep[]{pallanca+21b}, some residual differential reddening is still present in the CMD and made the derivation of the physical properties of the cluster particularly complex.
Nevertheless, by taking advantage of the differential reddening corrected CMD, we estimated a mean reddening $E(B-V)=6.33^{+0.05}_{-0.04}$ (confirming that Glimpse-C02 is one of most extincted GCs in our Galaxy), a distance $d=6.3^{+0.8}_{-0.3}$ kpc, and an absolute age $t=11.9^{+0.7}_{-0.6}$ Gyr.
We emphasize that this is the first age estimate of Glimpse-C02, made possible by the fact that the acquired data finally allowed us to identify the evolutionary sequences of the cluster, which revealed that it is a metal-rich and old GC. Its metallicity is consistent with that of 
bulge GCs and its distance from the Galactic center and the associated uncertainties ($\approx 2.6^{+0.6}_{-0.7}$ kpc) suggests it may be located closer to the bulge than previously reported. In the literature, it was placed in the inner transition region between the thin disk and the bulge \citep[see][]{kurtev+08}, while our estimate indicates a position somewhat nearer to the Galactic Center.
A kinematic analysis of the cluster’s motion within the MW could help clarifying whether belongs to the bulge or to the disk. In any case, given its metal-rich nature, we expect Glimpse-C02 to be an in-situ formed GC. 
Finally, we also studied the internal structure of the system exploiting the high-resolution of our dataset, first by estimating the center of gravity, which results to be offset by $3.12\arcsec$ from the literature one \citep[][]{harris+96}, then by building the radial density profile from star counts. This shows that Glimpse-C02 is a GC with a high concentration, which suggests an advanced stage of internal dynamical evolution. 
It is worth noticing, however, that the derived quantities are affected by large uncertainties, due to the high level of Galactic field contamination. The system's surface brightness profile allowed us to estimate its integrated absolute magnitude in the $H$ band ($M_{\rm H}=-7.9$) and its total mass ($M=3.57^{+0.22}_{-0.19}\times 10^4 M_{\odot}$), which collocate this stellar system in the low-mass end of the Galactic GCs mass distribution.

In summary, this analysis provided us with a very accurate characterization of the stellar population of Glimpse-C02 with a level of detail never reached before. This also demonstrates the great potential of high-resolution IR 
surveys to obtain a comprehensive view of the star cluster population in the MW, especially those located in its most extreme regions. Following these results, future prospects include obtaining longer-wavelength observations to improve the correction for differential reddening and, consequently, to better characterize the morphology of the evolutionary sequences. In addition, instruments such as JWST could provide higher-resolution images, which would help refine the differential reddening correction by resolving blended sources. They could also represent second epoch observations, allowing a solid distinction between cluster members and Galactic field interlopers via proper motion measurements and the determination of the cluster’s absolute motion in the plane of the sky. Combined with radial velocities from spectroscopy, these data could enable the reconstruction of the cluster’s orbit, and thus its dynamical history within the MW.

\begin{acknowledgements}
This work is part of the project {\it "GENESIS - Searching for the primordial structures of the Universe in the heart of the Galaxy"} (Advanced Grant FIS-2024-02056, PI:Ferraro), funded by the Italian MUR through the Fondo Italiano per la Scienza call. M.L. gratefully acknowledges funding from the European Union NextGenerationEU. ED acknowledges financial support from the INAF Data analysis Research Grant (PI E. Dalessandro) of the “Bando Astrofisica Fondamentale 2024”. DM acknowledges financial support from PRIN-MIUR-22: ``CHRONOS: adjusting the clock(s) to unveil the CHRONO-chemo-dynamical Structure of the Galaxy” (PI: S. Cassisi). 
\end{acknowledgements}

\bibliography{bib}

@ARTICLE{ferraro+26b, 
       author = {{Ferraro}, F.~R. and {Vesperini}, E. and {Lanzoni}, B. and {Romano}, D. and {Origlia}, L. and {Pallanca}, C. and {Fanelli}, C. and {Calura}, F. and {Dalessandro}, E. and {Massari}, D. and {Zullo}, G. and {Cadelano}, M.},
        title = "{Bulge Fossil Fragments as a new population of factories of gravitational wave sources in the Galaxy}",
      journal = {arXiv e-prints},
         year = 2026,
        month = mar,
          eid = {arXiv:2603.25127},
        pages = {arXiv:2603.25127},
archivePrefix = {arXiv},
       eprint = {2603.25127},
 primaryClass = {astro-ph.GA},
       adsurl = {https://ui.adsabs.harvard.edu/abs/2026arXiv260325127F}
}

@ARTICLE{ferraro+23,
       author = {{Ferraro}, Francesco R. and {Mucciarelli}, Alessio and {Lanzoni}, Barbara and {Pallanca}, Cristina and {Cadelano}, Mario and {Billi}, Alex and {Sills}, Alison and {Vesperini}, Enrico and {Dalessandro}, Emanuele and {Beccari}, Giacomo and {Monaco}, Lorenzo and {Mateo}, Mario},
        title = "{Fast rotating blue stragglers prefer loose clusters}",
      journal = {Nature Communications},
         year = 2023,
        month = may,
       volume = {14},
          eid = {2584},
        pages = {2584},
          doi = {10.1038/s41467-023-38153-w},
archivePrefix = {arXiv},
       eprint = {2305.08478},
 primaryClass = {astro-ph.SR},
       adsurl = {https://ui.adsabs.harvard.edu/abs/2023NatCo..14.2584F}
}

@ARTICLE{valenti10,
       author = {{Valenti}, E. and {Ferraro}, F.~R. and {Origlia}, L.},
        title = "{Near-infrared properties of 12 globular clusters towards the inner bulge of the Galaxy}",
      journal = {\mnras},
         year = 2010,
        month = mar,
       volume = {402},
       number = {3},
        pages = {1729-1739},
          doi = {10.1111/j.1365-2966.2009.15991.x},
archivePrefix = {arXiv},
       eprint = {0911.1264},
 primaryClass = {astro-ph.GA},
       adsurl = {https://ui.adsabs.harvard.edu/abs/2010MNRAS.402.1729V}
}

@ARTICLE{dalessandro08,
       author = {{Dalessandro}, E. and {Lanzoni}, B. and {Ferraro}, F.~R. and {Rood}, R.~T. and {Milone}, A. and {Piotto}, G. and {Valenti}, E.},
        title = "{Blue Straggler Stars in the Unusual Globular Cluster NGC 6388}",
      journal = {\apj},
         year = 2008,
        month = apr,
       volume = {677},
       number = {2},
        pages = {1069-1079},
          doi = {10.1086/529028},
archivePrefix = {arXiv},
       eprint = {0712.4272},
 primaryClass = {astro-ph},
       adsurl = {https://ui.adsabs.harvard.edu/abs/2008ApJ...677.1069D}
}

@ARTICLE{ferraro+26a,
       author = {{Ferraro}, Francesco R. and {Lanzoni}, Barbara and {Vesperini}, Enrico and {Dalessandro}, Emanuele and {Cadelano}, Mario and {Pallanca}, Cristina and {Beccari}, Giacomo and {Nardiello}, Domenico and {Libralato}, Mattia and {Piotto}, Giampaolo},
        title = "{A binary-related origin mediated by environmental conditions for blue straggler stars}",
      journal = {Nature Communications},
         year = 2026,
        month = jan,
       volume = {17},
       number = {1},
          eid = {768},
        pages = {768},
          doi = {10.1038/s41467-025-68159-5},
       adsurl = {https://ui.adsabs.harvard.edu/abs/2026NatCo..17..768F}
}

@ARTICLE{carretta+09,
       author = {{Carretta}, E. and {Bragaglia}, A. and {Gratton}, R. and {D'Orazi}, V. and {Lucatello}, S.},
        title = "{Intrinsic iron spread and a new metallicity scale for globular clusters}",
      journal = {\aap},
         year = 2009,
        month = dec,
       volume = {508},
       number = {2},
        pages = {695-706},
          doi = {10.1051/0004-6361/200913003},
archivePrefix = {arXiv},
       eprint = {0910.0675},
 primaryClass = {astro-ph.GA},
       adsurl = {https://ui.adsabs.harvard.edu/abs/2009A&A...508..695C}
}

@ARTICLE{libralato+22,
       author = {{Libralato}, Mattia and {Bellini}, Andrea and {Vesperini}, Enrico and {Piotto}, Giampaolo and {Milone}, Antonino P. and {van der Marel}, Roeland P. and {Anderson}, Jay and {Aparicio}, Antonio and {Barbuy}, Beatriz and {Bedin}, Luigi R. and {Borsato}, Luca and {Cassisi}, Santi and {Dalessandro}, Emanuele and {Ferraro}, Francesco R. and {King}, Ivan R. and {Lanzoni}, Barbara and {Nardiello}, Domenico and {Ortolani}, Sergio and {Sarajedini}, Ata and {Sohn}, Sangmo Tony},
        title = "{The Hubble Space Telescope UV Legacy Survey of Galactic Globular Clusters. XXIII. Proper-motion Catalogs and Internal Kinematics}",
      journal = {\apj},
         year = 2022,
        month = aug,
       volume = {934},
       number = {2},
          eid = {150},
        pages = {150},
          doi = {10.3847/1538-4357/ac7727},
archivePrefix = {arXiv},
       eprint = {2206.09924},
 primaryClass = {astro-ph.GA},
       adsurl = {https://ui.adsabs.harvard.edu/abs/2022ApJ...934..150L}
}

@ARTICLE{ferraro+18a,
       author = {{Ferraro}, F.~R. and {Mucciarelli}, A. and {Lanzoni}, B. and {Pallanca}, C. and {Lapenna}, E. and {Origlia}, L. and {Dalessandro}, E. and {Valenti}, E. and {Beccari}, G. and {Bellazzini}, M. and {Vesperini}, E. and {Varri}, A. and {Sollima}, A.},
        title = "{MIKiS: The Multi-instrument Kinematic Survey of Galactic Globular Clusters. I. Velocity Dispersion Profiles and Rotation Signals of 11 Globular Clusters}",
      journal = {\apj},
         year = 2018,
        month = jun,
       volume = {860},
       number = {1},
          eid = {50},
        pages = {50},
          doi = {10.3847/1538-4357/aabe2f},
archivePrefix = {arXiv},
       eprint = {1804.08618},
 primaryClass = {astro-ph.GA},
       adsurl = {https://ui.adsabs.harvard.edu/abs/2018ApJ...860...50F}
}

@ARTICLE{ferraro+18b,
       author = {{Ferraro}, F.~R. and {Lanzoni}, B. and {Raso}, S. and {Nardiello}, D. and {Dalessandro}, E. and {Vesperini}, E. and {Piotto}, G. and {Pallanca}, C. and {Beccari}, G. and {Bellini}, A. and {Libralato}, M. and {Anderson}, J. and {Aparicio}, A. and {Bedin}, L.~R. and {Cassisi}, S. and {Milone}, A.~P. and {Ortolani}, S. and {Renzini}, A. and {Salaris}, M. and {van der Marel}, R.~P.},
        title = "{The Hubble Space Telescope UV Legacy Survey of Galactic Globular Clusters. XV. The Dynamical Clock: Reading Cluster Dynamical Evolution from the Segregation Level of Blue Straggler Stars}",
      journal = {\apj},
         year = 2018,
        month = jun,
       volume = {860},
       number = {1},
          eid = {36},
        pages = {36},
          doi = {10.3847/1538-4357/aac01c},
archivePrefix = {arXiv},
       eprint = {1805.00968},
 primaryClass = {astro-ph.GA},
       adsurl = {https://ui.adsabs.harvard.edu/abs/2018ApJ...860...36F}
}

@ARTICLE{kamann+18,
       author = {{Kamann}, S. and {Husser}, T.-O. and {Dreizler}, S. and {Emsellem}, E. and {Weilbacher}, P.~M. and {Martens}, S. and {Bacon}, R. and {den Brok}, M. and {Giesers}, B. and {Krajnovi{\'c}}, D. and {Roth}, M.~M. and {Wendt}, M. and {Wisotzki}, L.},
        title = "{A stellar census in globular clusters with MUSE: The contribution of rotation to cluster dynamics studied with 200 000 stars}",
      journal = {\mnras},
         year = 2018,
        month = feb,
       volume = {473},
       number = {4},
        pages = {5591-5616},
          doi = {10.1093/mnras/stx2719},
archivePrefix = {arXiv},
       eprint = {1710.07257},
 primaryClass = {astro-ph.GA},
       adsurl = {https://ui.adsabs.harvard.edu/abs/2018MNRAS.473.5591K}
}

@ARTICLE{origlia+25,
       author = {{Origlia}, L. and {Ferraro}, F.~R. and {Fanelli}, C. and {Lanzoni}, B. and {Massari}, D. and {Dalessandro}, E. and {Pallanca}, C.},
        title = "{The manifest link between Terzan 5 and the Galactic bulge}",
      journal = {\aap},
         year = 2025,
        month = may,
       volume = {697},
          eid = {A19},
        pages = {A19},
          doi = {10.1051/0004-6361/202452110},
archivePrefix = {arXiv},
       eprint = {2503.17258},
 primaryClass = {astro-ph.GA},
       adsurl = {https://ui.adsabs.harvard.edu/abs/2025A&A...697A..19O}
}

@ARTICLE{ferraro+25,
       author = {{Ferraro}, F.~R. and {Chiappino}, L. and {Bartolomei}, A. and {Origlia}, L. and {Fanelli}, C. and {Lanzoni}, B. and {Pallanca}, C. and {Loriga}, M. and {Leanza}, S. and {Valenti}, E. and {Romano}, D. and {Mucciarelli}, A. and {Massari}, D. and {Cadelano}, M. and {Dalessandro}, E. and {Crociati}, C. and {Rich}, R.~M.},
        title = "{The Bulge Cluster Origin (BulCO) survey at the ESO-VLT: Probing the early history of the Milky Way assembly. Design and first results in Liller 1}",
      journal = {\aap},
         year = 2025,
        month = apr,
       volume = {696},
          eid = {A179},
        pages = {A179},
          doi = {10.1051/0004-6361/202554092},
archivePrefix = {arXiv},
       eprint = {2503.14642},
 primaryClass = {astro-ph.GA},
       adsurl = {https://ui.adsabs.harvard.edu/abs/2025A&A...696A.179F}
}

@ARTICLE{crociati+24,
       author = {{Crociati}, C. and {Cignoni}, M. and {Dalessandro}, E. and {Pallanca}, C. and {Massari}, D. and {Ferraro}, F.~R. and {Lanzoni}, B. and {Origlia}, L. and {Valenti}, E.},
        title = "{The star formation history of the first bulge fossil fragment candidate Terzan 5}",
      journal = {\aap},
         year = 2024,
        month = nov,
       volume = {691},
          eid = {A311},
        pages = {A311},
          doi = {10.1051/0004-6361/202451174},
archivePrefix = {arXiv},
       eprint = {2410.16971},
 primaryClass = {astro-ph.GA},
       adsurl = {https://ui.adsabs.harvard.edu/abs/2024A&A...691A.311C}
}

@ARTICLE{deimer+24,
       author = {{Alvarez Garay}, D.~A. and {Fanelli}, C. and {Origlia}, L. and {Pallanca}, C. and {Mucciarelli}, A. and {Chiappino}, L. and {Crociati}, C. and {Lanzoni}, B. and {Ferraro}, F.~R. and {Rich}, R.~M. and {Dalessandro}, E.},
        title = "{X-shooter spectroscopy of Liller 1 giant stars}",
      journal = {\aap},
         year = 2024,
        month = jun,
       volume = {686},
          eid = {A198},
        pages = {A198},
          doi = {10.1051/0004-6361/202449595},
archivePrefix = {arXiv},
       eprint = {2404.14130},
 primaryClass = {astro-ph.GA},
       adsurl = {https://ui.adsabs.harvard.edu/abs/2024A&A...686A.198A}
}

@ARTICLE{benjamin+03,
       author = {{Benjamin}, Robert A. and {Churchwell}, E. and {Babler}, Brian L. and {Bania}, T.~M. and {Clemens}, Dan P. and {Cohen}, Martin and {Dickey}, John M. and {Indebetouw}, R{\'e}my and {Jackson}, James M. and {Kobulnicky}, Henry A. and {Lazarian}, Alex and {Marston}, A.~P. and {Mathis}, John S. and {Meade}, Marilyn R. and {Seager}, Sara and {Stolovy}, S.~R. and {Watson}, C. and {Whitney}, Barbara A. and {Wolff}, Michael J. and {Wolfire}, Mark G.},
        title = "{GLIMPSE. I. An SIRTF Legacy Project to Map the Inner Galaxy}",
      journal = {\pasp},
         year = 2003,
        month = aug,
       volume = {115},
       number = {810},
        pages = {953-964},
          doi = {10.1086/376696},
archivePrefix = {arXiv},
       eprint = {astro-ph/0306274},
 primaryClass = {astro-ph},
       adsurl = {https://ui.adsabs.harvard.edu/abs/2003PASP..115..953B}
}

@ARTICLE{Cadelano2017,
	author = {Cadelano, M. and Pallanca, C. and Ferraro, F.R. and Dalessandro, E. and Lanzoni, B. and Patruno, A.},
	title = {The Optical Counterpart to the Accreting Millisecond X-Ray Pulsar SAX J1748.9-2021 in the Globular Cluster NGC 6440},
	year = {2017},
	journal = {Astrophysical Journal},
	volume = {844},
	number = {1},
	doi = {10.3847/1538-4357/aa7b7f},
	url = {https://www.scopus.com/inward/record.uri?eid=2-s2.0-85026354719&doi=10.3847%2f1538-4357%2faa7b7f&partnerID=40&md5=4ee461d1c13ef45bfc9481938a818371},
	type = {Article},
	publication_stage = {Final},
	source = {Scopus},
}

@ARTICLE{Cadelano2017_m71,
	author = {Cadelano, M. and Dalessandro, E. and Ferraro, F.R. and Miocchi, P. and Lanzoni, B. and Pallanca, C. and Massari, D.},
	title = {Proper Motions and Structural Parameters of the Galactic Globular Cluster M71},
	year = {2017},
	journal = {Astrophysical Journal},
	volume = {836},
	number = {2},
	doi = {10.3847/1538-4357/aa5ca5},
	url = {https://www.scopus.com/inward/record.uri?eid=2-s2.0-85014443014&doi=10.3847%2f1538-4357%2faa5ca5&partnerID=40&md5=20a48769d05aecf99e00fa2fa707c7a0},
	type = {Article},
	publication_stage = {Final},
	source = {Scopus},
}

@ARTICLE{Cadelano2022,
	author = {Cadelano, Mario and Ferraro, Francesco R. and Dalessandro, Emanuele and Lanzoni, Barbara and Pallanca, Cristina and Saracino, Sara},
	title = {Discovery of a Double Sequence of Blue Straggler Stars in the Core-collapsed Globular Cluster NGC 6256},
	year = {2022},
	journal = {Astrophysical Journal},
	volume = {941},
	number = {1},
	doi = {10.3847/1538-4357/aca016},
	url = {https://www.scopus.com/inward/record.uri?eid=2-s2.0-85144541150&doi=10.3847%2f1538-4357%2faca016&partnerID=40&md5=68d3808004a457111ff062411e24a705},
	type = {Article},
	publication_stage = {Final},
	source = {Scopus},
}

@ARTICLE{Beccari2023,
	author = {Beccari, G. and Cadelano, M. and Dalessandro, E.},
	title = {Dynamical state of the globular clusters Rup 106 and IC 4499},
	year = {2023},
	journal = {Astronomy and Astrophysics},
	volume = {670},
	doi = {10.1051/0004-6361/202244288},
	url = {https://www.scopus.com/inward/record.uri?eid=2-s2.0-85147141237&doi=10.1051%2f0004-6361%2f202244288&partnerID=40&md5=ce74bf5a50c544b0d289ee0171843acb},
	type = {Article},
	publication_stage = {Final},
	source = {Scopus},
}

@ARTICLE{Giusti2024,
	author = {Giusti, Camilla and Cadelano, Mario and Ferraro, Francesco R. and Lanzoni, Barbara and Pallanca, Cristina and Vesperini, Enrico and Dalessandro, Emanuele and Salaris, Maurizio},
	title = {Determining the dynamical age of the LMC globular cluster NGC 1835 using the 'dynamical clock': Star density profile and blue straggler stars},
	year = {2024},
	journal = {Astronomy and Astrophysics},
	volume = {687},
	doi = {10.1051/0004-6361/202450088},
	url = {https://www.scopus.com/inward/record.uri?eid=2-s2.0-85200128927&doi=10.1051%2f0004-6361%2f202450088&partnerID=40&md5=4ea45c6996f5b82d2950b94d19a1ae63},
	type = {Article},
	publication_stage = {Final},
	source = {Scopus},
}

@ARTICLE{Loriga2025,
	author = {Loriga, M. and Pallanca, C. and Ferraro, F.R. and Dalessandro, E. and Lanzoni, B. and Cadelano, M. and Origlia, L. and Fanelli, C. and Geisler, D. and Villanova, S.},
	title = {The bulge globular cluster Terzan 6 as seen from multi-conjugate adaptive optics and HST},
	year = {2025},
	journal = {Astronomy and Astrophysics},
	volume = {695},
	doi = {10.1051/0004-6361/202453482},
	url = {https://www.scopus.com/inward/record.uri?eid=2-s2.0-105000343605&doi=10.1051%2f0004-6361%2f202453482&partnerID=40&md5=05aecbc815bd652114fdf4fad205a800},
	type = {Article},
	publication_stage = {Final},
	source = {Scopus},
}

@ARTICLE{Cadelano2023,
	author = {Cadelano, Mario and Pallanca, Cristina and Dalessandro, Emanuele and Salaris, Maurizio and Mucciarelli, Alessio and Leanza, Silvia and Ferraro, Francesco R. and Lanzoni, Barbara and Rosie Chen, C.-H. and Freire, Paulo C. C. and Heinke, Craig and Ransom, Scott M.},
	title = {JWST uncovers helium and water abundance variations in the bulge globular cluster NGC 6440},
	year = {2023},
	journal = {Astronomy and Astrophysics},
	volume = {679},
	doi = {10.1051/0004-6361/202347961},
	url = {https://www.scopus.com/inward/record.uri?eid=2-s2.0-85179123846&doi=10.1051%2f0004-6361%2f202347961&partnerID=40&md5=59c1915f397b2e3b9b6e004eeda2839f},
	type = {Article},
	publication_stage = {Final},
	source = {Scopus},
}

@ARTICLE{Dalessandro2019,
	author = {Dalessandro, E. and Ferraro, F.R. and Bastian, N. and Cadelano, M. and Lanzoni, B. and Raso, S.},
	title = {The double blue-straggler sequence in NGC 2173: An artifact of field contamination},
	year = {2019},
	journal = {Astronomy and Astrophysics},
	volume = {621},
	doi = {10.1051/0004-6361/201834011},
	url = {https://www.scopus.com/inward/record.uri?eid=2-s2.0-85059630366&doi=10.1051%2f0004-6361%2f201834011&partnerID=40&md5=dbb6cf33b49edf3b6eac74fad6c62fd9}
}

@ARTICLE{Giusti2023_kmk,
	author = {Giusti, Camilla and Cadelano, Mario and Ferraro, Francesco R. and Lanzoni, Barbara and Leanza, Silvia and Pallanca, Cristina and Vesperini, Enrico and Dalessandro, Emanuele and Mucciarelli, Alessio},
	title = {An Ongoing Tidal Capture in the Large Magellanic Cloud: The Low-mass Star Cluster KMK 88-10 Captured by the Massive Globular Cluster NGC 1835?},
	year = {2023},
	journal = {Astrophysical Journal},
	volume = {953},
	number = {2},
	doi = {10.3847/1538-4357/ace18e},
	url = {https://www.scopus.com/inward/record.uri?eid=2-s2.0-85167876520&doi=10.3847%2f1538-4357%2face18e&partnerID=40&md5=23a6eb3dbea35338014c9c9d1dc198ac}
}

@ARTICLE{borissova+14,
       author = {{Borissova}, J. and {Chen{\'e}}, A. -N. and {Ram{\'\i}rez Alegr{\'\i}a}, S. and {Sharma}, S. and {Clarke}, J.~R.~A. and {Kurtev}, R. and {Negueruela}, I. and {Marco}, A. and {Amigo}, P. and {Minniti}, D. and {Bica}, E. and {Bonatto}, C. and {Catelan}, M. and {Fierro}, C. and {Geisler}, D. and {Gromadzki}, M. and {Hempel}, M. and {Hanson}, M.~M. and {Ivanov}, V.~D. and {Lucas}, P. and {Majaess}, D. and {Moni Bidin}, C. and {Popescu}, B. and {Saito}, R.~K.},
        title = "{New galactic star clusters discovered in the VVV survey. Candidates projected on the inner disk and bulge}",
      journal = {\aap},
         year = 2014,
        month = sep,
       volume = {569},
          eid = {A24},
        pages = {A24},
          doi = {10.1051/0004-6361/201322483},
archivePrefix = {arXiv},
       eprint = {1406.7051},
 primaryClass = {astro-ph.GA},
       adsurl = {https://ui.adsabs.harvard.edu/abs/2014A&A...569A..24B}
}

@ARTICLE{cadelano+20,
       author = {{Cadelano}, M. and {Saracino}, S. and {Dalessandro}, E. and {Ferraro}, F.~R. and {Lanzoni}, B. and {Massari}, D. and {Pallanca}, C. and {Salaris}, M.},
        title = "{Digging for Relics of the Past: The Ancient and Obscured Bulge Globular Cluster NGC 6256}",
      journal = {\apj},
         year = 2020,
        month = may,
       volume = {895},
       number = {1},
          eid = {54},
        pages = {54},
          doi = {10.3847/1538-4357/ab88b3},
archivePrefix = {arXiv},
       eprint = {2004.06131},
 primaryClass = {astro-ph.GA},
       adsurl = {https://ui.adsabs.harvard.edu/abs/2020ApJ...895...54C}
}

@ARTICLE{camargo+18,
       author = {{Camargo}, Denilso},
        title = "{Five New Globular Clusters Discovered in the Galactic Bulge}",
      journal = {\apjl},
         year = 2018,
        month = jun,
       volume = {860},
       number = {2},
          eid = {L27},
        pages = {L27},
          doi = {10.3847/2041-8213/aacc68},
       adsurl = {https://ui.adsabs.harvard.edu/abs/2018ApJ...860L..27C}
}

@ARTICLE{camargo+19,
       author = {{Camargo}, D. and {Minniti}, D.},
        title = "{Three candidate globular clusters discovered in the Galactic bulge}",
      journal = {\mnras},
         year = 2019,
        month = mar,
       volume = {484},
       number = {1},
        pages = {L90-L94},
          doi = {10.1093/mnrasl/slz010},
archivePrefix = {arXiv},
       eprint = {1901.08574},
 primaryClass = {astro-ph.GA},
       adsurl = {https://ui.adsabs.harvard.edu/abs/2019MNRAS.484L..90C}
}

@ARTICLE{cardelli+89,
       author = {{Cardelli}, Jason A. and {Clayton}, Geoffrey C. and {Mathis}, John S.},
        title = "{The Relationship between Infrared, Optical, and Ultraviolet Extinction}",
      journal = {\apj},
         year = 1989,
        month = oct,
       volume = {345},
        pages = {245},
          doi = {10.1086/167900},
       adsurl = {https://ui.adsabs.harvard.edu/abs/1989ApJ...345..245C}
}

@ARTICLE{crociati+23,
       author = {{Crociati}, Chiara and {Valenti}, Elena and {Ferraro}, Francesco R. and {Pallanca}, Cristina and {Lanzoni}, Barbara and {Cadelano}, Mario and {Fanelli}, Cristiano and {Origlia}, Livia and {Leanza}, Silvia and {Dalessandro}, Emanuele and {Mucciarelli}, Alessio and {Rich}, R. Michael},
        title = "{First Evidence of Multi-iron Subpopulations in the Bulge Fossil Fragment Candidate Liller 1}",
      journal = {\apj},
         year = 2023,
        month = jul,
       volume = {951},
       number = {1},
          eid = {17},
        pages = {17},
          doi = {10.3847/1538-4357/acd382},
archivePrefix = {arXiv},
       eprint = {2305.04595},
 primaryClass = {astro-ph.GA},
       adsurl = {https://ui.adsabs.harvard.edu/abs/2023ApJ...951...17C}
}

@ARTICLE{dalessandro+22,
       author = {{Dalessandro}, Emanuele and {Crociati}, Chiara and {Cignoni}, Michele and {Ferraro}, Francesco R. and {Lanzoni}, Barbara and {Origlia}, Livia and {Pallanca}, Cristina and {Rich}, R. Michael and {Saracino}, Sara and {Valenti}, Elena},
        title = "{Clues to the Formation of Liller 1 from Modeling Its Complex Star Formation History}",
      journal = {\apj},
         year = 2022,
        month = dec,
       volume = {940},
       number = {2},
          eid = {170},
        pages = {170},
          doi = {10.3847/1538-4357/ac9907},
archivePrefix = {arXiv},
       eprint = {2210.05694},
 primaryClass = {astro-ph.GA},
       adsurl = {https://ui.adsabs.harvard.edu/abs/2022ApJ...940..170D}
}

@ARTICLE{dias+22,
       author = {{Dias}, B. and {Palma}, T. and {Minniti}, D. and {Fern{\'a}ndez-Trincado}, J.~G. and {Alonso-Garc{\'\i}a}, J. and {Barbuy}, B. and {Clari{\'a}}, J.~J. and {Gomez}, M. and {Saito}, R.~K.},
        title = "{FSR 1776: A new globular cluster in the Galactic bulge?}",
      journal = {\aap},
         year = 2022,
        month = jan,
       volume = {657},
          eid = {A67},
        pages = {A67},
          doi = {10.1051/0004-6361/202141580},
archivePrefix = {arXiv},
       eprint = {2110.00868},
 primaryClass = {astro-ph.GA},
       adsurl = {https://ui.adsabs.harvard.edu/abs/2022A&A...657A..67D}
}

@ARTICLE{dolphin+2000,
       author = {{Dolphin}, Andrew E.},
        title = "{WFPC2 Stellar Photometry with HSTPHOT}",
      journal = {\pasp},
         year = 2000,
        month = oct,
       volume = {112},
       number = {776},
        pages = {1383-1396},
          doi = {10.1086/316630},
archivePrefix = {arXiv},
       eprint = {astro-ph/0006217},
 primaryClass = {astro-ph},
       adsurl = {https://ui.adsabs.harvard.edu/abs/2000PASP..112.1383D}
}

@software{dolphin+16,
       author = {{Dolphin}, Andrew},
        title = "{DOLPHOT: Stellar photometry}",
 howpublished = {Astrophysics Source Code Library, record ascl:1608.013},
         year = 2016,
        month = aug,
          eid = {ascl:1608.013},
archivePrefix = {ascl},
       eprint = {1608.013},
       adsurl = {https://ui.adsabs.harvard.edu/abs/2016ascl.soft08013D}
}

@ARTICLE{fanelli+24,
       author = {{Fanelli}, C. and {Origlia}, L. and {Mucciarelli}, A. and {Ferraro}, F.~R. and {Rich}, R.~M. and {Lanzoni}, B. and {Massari}, D. and {Pallanca}, C. and {Dalessandro}, E. and {Loriga}, M.},
        title = "{Detailed chemical abundances of the globular cluster Terzan 6 in the inner bulge}",
      journal = {\aap},
         year = 2024,
        month = aug,
       volume = {688},
          eid = {A154},
        pages = {A154},
          doi = {10.1051/0004-6361/202450556},
archivePrefix = {arXiv},
       eprint = {2406.07180},
 primaryClass = {astro-ph.GA},
       adsurl = {https://ui.adsabs.harvard.edu/abs/2024A&A...688A.154F}
}

@ARTICLE{ferraro+09,
       author = {{Ferraro}, F.~R. and {Dalessandro}, E. and {Mucciarelli}, A. and {Beccari}, G. and {Rich}, R.~M. and {Origlia}, L. and {Lanzoni}, B. and {Rood}, R.~T. and {Valenti}, E. and {Bellazzini}, M. and {Ransom}, S.~M. and {Cocozza}, G.},
        title = "{The cluster Terzan 5 as a remnant of a primordial building block of the Galactic bulge}",
      journal = {\nat},
         year = 2009,
        month = nov,
       volume = {462},
       number = {7272},
        pages = {483-486},
          doi = {10.1038/nature08581},
archivePrefix = {arXiv},
       eprint = {0912.0192},
 primaryClass = {astro-ph.GA},
       adsurl = {https://ui.adsabs.harvard.edu/abs/2009Natur.462..483F}
}

@ARTICLE{ferraro+16,
       author = {{Ferraro}, F.~R. and {Massari}, D. and {Dalessandro}, E. and {Lanzoni}, B. and {Origlia}, L. and {Rich}, R.~M. and {Mucciarelli}, A.},
        title = "{The Age of the Young Bulge-like Population in the Stellar System Terzan 5: Linking the Galactic Bulge to the High-z Universe}",
      journal = {\apj},
         year = 2016,
        month = sep,
       volume = {828},
       number = {2},
          eid = {75},
        pages = {75},
          doi = {10.3847/0004-637X/828/2/75},
archivePrefix = {arXiv},
       eprint = {1609.01515},
 primaryClass = {astro-ph.GA},
       adsurl = {https://ui.adsabs.harvard.edu/abs/2016ApJ...828...75F}
}

@ARTICLE{ferraro+21,
       author = {{Ferraro}, F.~R. and {Pallanca}, C. and {Lanzoni}, B. and {Crociati}, C. and {Dalessandro}, E. and {Origlia}, L. and {Rich}, R.~M. and {Saracino}, S. and {Mucciarelli}, A. and {Valenti}, E. and {Geisler}, D. and {Mauro}, F. and {Villanova}, S. and {Moni Bidin}, C. and {Beccari}, G.},
        title = "{A new class of fossil fragments from the hierarchical assembly of the Galactic bulge}",
      journal = {Nature Astronomy},
         year = 2021,
        month = jan,
       volume = {5},
        pages = {311-318},
          doi = {10.1038/s41550-020-01267-y},
archivePrefix = {arXiv},
       eprint = {2011.09966},
 primaryClass = {astro-ph.GA},
       adsurl = {https://ui.adsabs.harvard.edu/abs/2021NatAs...5..311F}
}

@ARTICLE{fitzpatrick+90,
       author = {{Fitzpatrick}, Edward L. and {Massa}, Derck},
        title = "{An Analysis of the Shapes of Ultraviolet Extinction Curves. III. an Atlas of Ultraviolet Extinction Curves}",
      journal = {\apjs},
         year = 1990,
        month = jan,
       volume = {72},
        pages = {163},
          doi = {10.1086/191413},
       adsurl = {https://ui.adsabs.harvard.edu/abs/1990ApJS...72..163F}
}

@ARTICLE{fitzpatrick+99,
       author = {{Fitzpatrick}, Edward L.},
        title = "{Correcting for the Effects of Interstellar Extinction}",
      journal = {\pasp},
         year = 1999,
        month = jan,
       volume = {111},
       number = {755},
        pages = {63-75},
          doi = {10.1086/316293},
archivePrefix = {arXiv},
       eprint = {astro-ph/9809387},
 primaryClass = {astro-ph},
       adsurl = {https://ui.adsabs.harvard.edu/abs/1999PASP..111...63F}
}

@ARTICLE{froebrich+07,
       author = {{Froebrich}, D. and {Meusinger}, H. and {Scholz}, A.},
        title = "{FSR1735 - a new globular cluster candidate in the inner Galaxy}",
      journal = {\mnras},
         year = 2007,
        month = may,
       volume = {377},
       number = {1},
        pages = {L54-L58},
          doi = {10.1111/j.1745-3933.2007.00302.x},
archivePrefix = {arXiv},
       eprint = {astro-ph/0703318},
 primaryClass = {astro-ph},
       adsurl = {https://ui.adsabs.harvard.edu/abs/2007MNRAS.377L..54F}
}

@ARTICLE{gaia+23,
       author = {{Gaia Collaboration} and {Vallenari}, A. and {Brown}, A.~G.~A. and {Prusti}, T. and {de Bruijne}, J.~H.~J. and {Arenou}, F. and {Babusiaux}, C. and {Biermann}, M. and {Creevey}, O.~L. and {Ducourant}, C. and {Evans}, D.~W. and {Eyer}, L. and {Guerra}, R. and {Hutton}, A. and {Jordi}, C. and {Klioner}, S.~A. and {Lammers}, U.~L. and {Lindegren}, L. and {Luri}, X. and {Mignard}, F. and {Panem}, C. and {Pourbaix}, D. and {Randich}, S. and {Sartoretti}, P. and {Soubiran}, C. and {Tanga}, P. and {Walton}, N.~A. and {Bailer-Jones}, C.~A.~L. and {Bastian}, U. and {Drimmel}, R. and {Jansen}, F. and {Katz}, D. and {Lattanzi}, M.~G. and {van Leeuwen}, F. and {Bakker}, J. and {Cacciari}, C. and {Casta{\~n}eda}, J. and {De Angeli}, F. and {Fabricius}, C. and {Fouesneau}, M. and {Fr{\'e}mat}, Y. and {Galluccio}, L. and {Guerrier}, A. and {Heiter}, U. and {Masana}, E. and {Messineo}, R. and {Mowlavi}, N. and {Nicolas}, C. and {Nienartowicz}, K. and {Pailler}, F. and {Panuzzo}, P. and {Riclet}, F. and {Roux}, W. and {Seabroke}, G.~M. and {Sordo}, R. and {Th{\'e}venin}, F. and {Gracia-Abril}, G. and {Portell}, J. and {Teyssier}, D. and {Altmann}, M. and {Andrae}, R. and {Audard}, M. and {Bellas-Velidis}, I. and {Benson}, K. and {Berthier}, J. and {Blomme}, R. and {Burgess}, P.~W. and {Busonero}, D. and {Busso}, G. and {C{\'a}novas}, H. and {Carry}, B. and {Cellino}, A. and {Cheek}, N. and {Clementini}, G. and {Damerdji}, Y. and {Davidson}, M. and {de Teodoro}, P. and {Nu{\~n}ez Campos}, M. and {Delchambre}, L. and {Dell'Oro}, A. and {Esquej}, P. and {Fern{\'a}ndez-Hern{\'a}ndez}, J. and {Fraile}, E. and {Garabato}, D. and {Garc{\'\i}a-Lario}, P. and {Gosset}, E. and {Haigron}, R. and {Halbwachs}, J. -L. and {Hambly}, N.~C. and {Harrison}, D.~L. and {Hern{\'a}ndez}, J. and {Hestroffer}, D. and {Hodgkin}, S.~T. and {Holl}, B. and {Jan{\ss}en}, K. and {Jevardat de Fombelle}, G. and {Jordan}, S. and {Krone-Martins}, A. and {Lanzafame}, A.~C. and {L{\"o}ffler}, W. and {Marchal}, O. and {Marrese}, P.~M. and {Moitinho}, A. and {Muinonen}, K. and {Osborne}, P. and {Pancino}, E. and {Pauwels}, T. and {Recio-Blanco}, A. and {Reyl{\'e}}, C. and {Riello}, M. and {Rimoldini}, L. and {Roegiers}, T. and {Rybizki}, J. and {Sarro}, L.~M. and {Siopis}, C. and {Smith}, M. and {Sozzetti}, A. and {Utrilla}, E. and {van Leeuwen}, M. and {Abbas}, U. and {{\'A}brah{\'a}m}, P. and {Abreu Aramburu}, A. and {Aerts}, C. and {Aguado}, J.~J. and {Ajaj}, M. and {Aldea-Montero}, F. and {Altavilla}, G. and {{\'A}lvarez}, M.~A. and {Alves}, J. and {Anders}, F. and {Anderson}, R.~I. and {Anglada Varela}, E. and {Antoja}, T. and {Baines}, D. and {Baker}, S.~G. and {Balaguer-N{\'u}{\~n}ez}, L. and {Balbinot}, E. and {Balog}, Z. and {Barache}, C. and {Barbato}, D. and {Barros}, M. and {Barstow}, M.~A. and {Bartolom{\'e}}, S. and {Bassilana}, J. -L. and {Bauchet}, N. and {Becciani}, U. and {Bellazzini}, M. and {Berihuete}, A. and {Bernet}, M. and {Bertone}, S. and {Bianchi}, L. and {Binnenfeld}, A. and {Blanco-Cuaresma}, S. and {Blazere}, A. and {Boch}, T. and {Bombrun}, A. and {Bossini}, D. and {Bouquillon}, S. and {Bragaglia}, A. and {Bramante}, L. and {Breedt}, E. and {Bressan}, A. and {Brouillet}, N. and {Brugaletta}, E. and {Bucciarelli}, B. and {Burlacu}, A. and {Butkevich}, A.~G. and {Buzzi}, R. and {Caffau}, E. and {Cancelliere}, R. and {Cantat-Gaudin}, T. and {Carballo}, R. and {Carlucci}, T. and {Carnerero}, M.~I. and {Carrasco}, J.~M. and {Casamiquela}, L. and {Castellani}, M. and {Castro-Ginard}, A. and {Chaoul}, L. and {Charlot}, P. and {Chemin}, L. and {Chiaramida}, V. and {Chiavassa}, A. and {Chornay}, N. and {Comoretto}, G. and {Contursi}, G. and {Cooper}, W.~J. and {Cornez}, T. and {Cowell}, S. and {Crifo}, F. and {Cropper}, M. and {Crosta}, M. and {Crowley}, C. and {Dafonte}, C. and {Dapergolas}, A. and {David}, M. and {David}, P. and {de Laverny}, P. and {De Luise}, F. and {De March}, R.},
        title = "{Gaia Data Release 3. Summary of the content and survey properties}",
      journal = {\aap},
         year = 2023,
        month = jun,
       volume = {674},
          eid = {A1},
        pages = {A1},
          doi = {10.1051/0004-6361/202243940},
archivePrefix = {arXiv},
       eprint = {2208.00211},
 primaryClass = {astro-ph.GA},
       adsurl = {https://ui.adsabs.harvard.edu/abs/2023A&A...674A...1G}
}

@ARTICLE{garro+20,
       author = {{Garro}, E.~R. and {Minniti}, D. and {G{\'o}mez}, M. and {Alonso-Garc{\'\i}a}, J. and {Barb{\'a}}, R.~H. and {Barbuy}, B. and {Clari{\'a}}, J.~J. and {Chen{\'e}}, A.~N. and {Dias}, B. and {Hempel}, M. and {Ivanov}, V.~D. and {Lucas}, P.~W. and {Majaess}, D. and {Mauro}, F. and {Moni Bidin}, C. and {Palma}, T. and {Pullen}, J.~B. and {Saito}, R.~K. and {Smith}, L. and {Surot}, F. and {Ram{\'\i}rez Alegr{\'\i}a}, S. and {Rejkuba}, M. and {Ripepi}, V. and {Fern{\'a}ndez Trincado}, J.},
        title = "{VVVX-Gaia discovery of a low luminosity globular cluster in the Milky Way disk}",
      journal = {\aap},
         year = 2020,
        month = oct,
       volume = {642},
          eid = {L19},
        pages = {L19},
          doi = {10.1051/0004-6361/202039233},
archivePrefix = {arXiv},
       eprint = {2010.02113},
 primaryClass = {astro-ph.GA},
       adsurl = {https://ui.adsabs.harvard.edu/abs/2020A&A...642L..19G}
}

@ARTICLE{garro+21,
       author = {{Garro}, E.~R. and {Minniti}, D. and {G{\'o}mez}, M. and {Alonso-Garc{\'\i}a}, J. and {Palma}, T. and {Smith}, L.~C. and {Ripepi}, V.},
        title = "{Confirmation and physical characterization of the new bulge globular cluster Patchick 99 from the VVV and Gaia surveys}",
      journal = {\aap},
         year = 2021,
        month = may,
       volume = {649},
          eid = {A86},
        pages = {A86},
          doi = {10.1051/0004-6361/202039255},
archivePrefix = {arXiv},
       eprint = {2103.03592},
 primaryClass = {astro-ph.GA},
       adsurl = {https://ui.adsabs.harvard.edu/abs/2021A&A...649A..86G}
}

@ARTICLE{garro+22,
       author = {{Garro}, E.~R. and {Minniti}, D. and {G{\'o}mez}, M. and {Alonso-Garc{\'\i}a}, J. and {Ripepi}, V. and {Fern{\'a}ndez-Trincado}, J.~G. and {Vivanco C{\'a}diz}, F.},
        title = "{Inspection of 19 globular cluster candidates in the Galactic bulge with the VVV survey}",
      journal = {\aap},
         year = 2022,
        month = feb,
       volume = {658},
          eid = {A120},
        pages = {A120},
          doi = {10.1051/0004-6361/202141819},
archivePrefix = {arXiv},
       eprint = {2111.08317},
 primaryClass = {astro-ph.GA},
       adsurl = {https://ui.adsabs.harvard.edu/abs/2022A&A...658A.120G}
}

@ARTICLE{harris+96,
       author = {{Harris}, William E.},
        title = "{A Catalog of Parameters for Globular Clusters in the Milky Way}",
      journal = {\aj},
         year = 1996,
        month = oct,
       volume = {112},
        pages = {1487},
          doi = {10.1086/118116},
       adsurl = {https://ui.adsabs.harvard.edu/abs/1996AJ....112.1487H}
}

@ARTICLE{king+66,
       author = {{King}, Ivan R.},
        title = "{The structure of star clusters. III. Some simple dynamical models}",
      journal = {\aj},
         year = 1966,
        month = feb,
       volume = {71},
        pages = {64},
          doi = {10.1086/109857},
       adsurl = {https://ui.adsabs.harvard.edu/abs/1966AJ.....71...64K}
}

@ARTICLE{kurtev+08,
       author = {{Kurtev}, R. and {Ivanov}, V.~D. and {Borissova}, J. and {Ortolani}, S.},
        title = "{Obscured clusters. II. GLIMPSE-C02 - A new metal rich globular cluster in the Milky Way}",
      journal = {\aap},
         year = 2008,
        month = oct,
       volume = {489},
       number = {2},
        pages = {583-587},
          doi = {10.1051/0004-6361:200809425},
archivePrefix = {arXiv},
       eprint = {0808.1565},
 primaryClass = {astro-ph},
       adsurl = {https://ui.adsabs.harvard.edu/abs/2008A&A...489..583K}
}

@ARTICLE{lanzoni+10,
       author = {{Lanzoni}, B. and {Ferraro}, F.~R. and {Dalessandro}, E. and {Mucciarelli}, A. and {Beccari}, G. and {Miocchi}, P. and {Bellazzini}, M. and {Rich}, R.~M. and {Origlia}, L. and {Valenti}, E. and {Rood}, R.~T. and {Ransom}, S.~M.},
        title = "{New Density Profile and Structural Parameters of the Complex Stellar System Terzan 5}",
      journal = {\apj},
         year = 2010,
        month = jul,
       volume = {717},
       number = {2},
        pages = {653-657},
          doi = {10.1088/0004-637X/717/2/653},
archivePrefix = {arXiv},
       eprint = {1005.2847},
 primaryClass = {astro-ph.GA},
       adsurl = {https://ui.adsabs.harvard.edu/abs/2010ApJ...717..653L}
}

@ARTICLE{marin+09,
       author = {{Mar{\'\i}n-Franch}, Antonio and {Aparicio}, Antonio and {Piotto}, Giampaolo and {Rosenberg}, Alfred and {Chaboyer}, Brian and {Sarajedini}, Ata and {Siegel}, Michael and {Anderson}, Jay and {Bedin}, Luigi R. and {Dotter}, Aaron and {Hempel}, Maren and {King}, Ivan and {Majewski}, Steven and {Milone}, Antonino P. and {Paust}, Nathaniel and {Reid}, I. Neill},
        title = "{The ACS Survey of Galactic Globular Clusters. VII. Relative Ages}",
      journal = {\apj},
         year = 2009,
        month = apr,
       volume = {694},
       number = {2},
        pages = {1498-1516},
          doi = {10.1088/0004-637X/694/2/1498},
archivePrefix = {arXiv},
       eprint = {0812.4541},
 primaryClass = {astro-ph},
       adsurl = {https://ui.adsabs.harvard.edu/abs/2009ApJ...694.1498M}
}

@ARTICLE{massari+14,
       author = {{Massari}, D. and {Mucciarelli}, A. and {Ferraro}, F.~R. and {Origlia}, L. and {Rich}, R.~M. and {Lanzoni}, B. and {Dalessandro}, E. and {Valenti}, E. and {Ibata}, R. and {Lovisi}, L. and {Bellazzini}, M. and {Reitzel}, D.},
        title = "{Ceci N'est Pas a Globular Cluster: The Metallicity Distribution of the Stellar System Terzan 5}",
      journal = {\apj},
         year = 2014,
        month = nov,
       volume = {795},
       number = {1},
          eid = {22},
        pages = {22},
          doi = {10.1088/0004-637X/795/1/22},
archivePrefix = {arXiv},
       eprint = {1409.1682},
 primaryClass = {astro-ph.SR},
       adsurl = {https://ui.adsabs.harvard.edu/abs/2014ApJ...795...22M}
}

@ARTICLE{mercer+05,
       author = {{Mercer}, E.~P. and {Clemens}, D.~P. and {Meade}, M.~R. and {Babler}, B.~L. and {Indebetouw}, R. and {Whitney}, B.~A. and {Watson}, C. and {Wolfire}, M.~G. and {Wolff}, M.~J. and {Bania}, T.~M. and {Benjamin}, R.~A. and {Cohen}, M. and {Dickey}, J.~M. and {Jackson}, J.~M. and {Kobulnicky}, H.~A. and {Mathis}, J.~S. and {Stauffer}, J.~R. and {Stolovy}, S.~R. and {Uzpen}, B. and {Churchwell}, E.~B.},
        title = "{New Star Clusters Discovered in the GLIMPSE Survey}",
      journal = {\apj},
         year = 2005,
        month = dec,
       volume = {635},
       number = {1},
        pages = {560-569},
          doi = {10.1086/497260},
       adsurl = {https://ui.adsabs.harvard.edu/abs/2005ApJ...635..560M}
}

@ARTICLE{minniti+10,
       author = {{Minniti}, D. and {Lucas}, P.~W. and {Emerson}, J.~P. and {Saito}, R.~K. and {Hempel}, M. and {Pietrukowicz}, P. and {Ahumada}, A.~V. and {Alonso}, M.~V. and {Alonso-Garcia}, J. and {Arias}, J.~I. and {Bandyopadhyay}, R.~M. and {Barb{\'a}}, R.~H. and {Barbuy}, B. and {Bedin}, L.~R. and {Bica}, E. and {Borissova}, J. and {Bronfman}, L. and {Carraro}, G. and {Catelan}, M. and {Clari{\'a}}, J.~J. and {Cross}, N. and {de Grijs}, R. and {D{\'e}k{\'a}ny}, I. and {Drew}, J.~E. and {Fari{\~n}a}, C. and {Feinstein}, C. and {Fern{\'a}ndez Laj{\'u}s}, E. and {Gamen}, R.~C. and {Geisler}, D. and {Gieren}, W. and {Goldman}, B. and {Gonzalez}, O.~A. and {Gunthardt}, G. and {Gurovich}, S. and {Hambly}, N.~C. and {Irwin}, M.~J. and {Ivanov}, V.~D. and {Jord{\'a}n}, A. and {Kerins}, E. and {Kinemuchi}, K. and {Kurtev}, R. and {L{\'o}pez-Corredoira}, M. and {Maccarone}, T. and {Masetti}, N. and {Merlo}, D. and {Messineo}, M. and {Mirabel}, I.~F. and {Monaco}, L. and {Morelli}, L. and {Padilla}, N. and {Palma}, T. and {Parisi}, M.~C. and {Pignata}, G. and {Rejkuba}, M. and {Roman-Lopes}, A. and {Sale}, S.~E. and {Schreiber}, M.~R. and {Schr{\"o}der}, A.~C. and {Smith}, M. and {Sodr{\'e}}, Jr., L. and {Soto}, M. and {Tamura}, M. and {Tappert}, C. and {Thompson}, M.~A. and {Toledo}, I. and {Zoccali}, M. and {Pietrzynski}, G.},
        title = "{VISTA Variables in the Via Lactea (VVV): The public ESO near-IR variability survey of the Milky Way}",
      journal = {\na},
         year = 2010,
        month = jul,
       volume = {15},
       number = {5},
        pages = {433-443},
          doi = {10.1016/j.newast.2009.12.002},
archivePrefix = {arXiv},
       eprint = {0912.1056},
 primaryClass = {astro-ph.GA},
       adsurl = {https://ui.adsabs.harvard.edu/abs/2010NewA...15..433M}
}

@ARTICLE{minniti+17,
       author = {{Minniti}, Dante and {Palma}, Tali and {D{\'e}k{\'a}ny}, Istvan and {Hempel}, Maren and {Rejkuba}, Marina and {Pullen}, Joyce and {Alonso-Garc{\'\i}a}, Javier and {Barb{\'a}}, Rodolfo and {Barbuy}, Beatriz and {Bica}, Eduardo and {Bonatto}, Charles and {Borissova}, Jura and {Catelan}, Marcio and {Carballo-Bello}, Julio A. and {Chene}, Andre Nicolas and {Clari{\'a}}, Juan Jos{\'e} and {Cohen}, Roger E. and {Contreras Ramos}, Rodrigo and {Dias}, Bruno and {Emerson}, Jim and {Froebrich}, Dirk and {Buckner}, Anne S.~M. and {Geisler}, Douglas and {Gonzalez}, Oscar A. and {Gran}, Felipe and {Hajdu}, Gergely and {Irwin}, Mike and {Ivanov}, Valentin D. and {Kurtev}, Radostin and {Lucas}, Philip W. and {Majaess}, Daniel and {Mauro}, Francesco and {Moni-Bidin}, Christian and {Navarrete}, Camila and {Ram{\'\i}rez Alegr{\'\i}a}, Sebastian and {Saito}, Roberto K. and {Valenti}, Elena and {Zoccali}, Manuela},
        title = "{FSR 1716: A New Milky Way Globular Cluster Confirmed Using VVV RR Lyrae Stars}",
      journal = {\apjl},
         year = 2017,
        month = mar,
       volume = {838},
       number = {1},
          eid = {L14},
        pages = {L14},
          doi = {10.3847/2041-8213/838/1/L14},
archivePrefix = {arXiv},
       eprint = {1703.02033},
 primaryClass = {astro-ph.GA},
       adsurl = {https://ui.adsabs.harvard.edu/abs/2017ApJ...838L..14M}
}

@ARTICLE{minniti+21a,
       author = {{Minniti}, Dante and {Fern{\'a}ndez-Trincado}, Jos{\'e} G. and {G{\'o}mez}, Mat{\'\i}as and {Smith}, Leigh C. and {Lucas}, Philip W. and {Contreras Ramos}, Rodrigo},
        title = "{Discovery of a new nearby globular cluster with extreme kinematics located in the extension of a halo stream}",
      journal = {\aap},
         year = 2021,
        month = jun,
       volume = {650},
          eid = {L11},
        pages = {L11},
          doi = {10.1051/0004-6361/202141129},
archivePrefix = {arXiv},
       eprint = {2106.01383},
 primaryClass = {astro-ph.GA},
       adsurl = {https://ui.adsabs.harvard.edu/abs/2021A&A...650L..11M}
}

@ARTICLE{minniti+21b,
       author = {{Minniti}, D. and {Palma}, T. and {Camargo}, D. and {Chijani-Saballa}, M. and {Alonso-Garc{\'\i}a}, J. and {Clari{\'a}}, J.~J. and {Dias}, B. and {G{\'o}mez}, M. and {Pullen}, J.~B. and {Saito}, R.~K.},
        title = "{An intriguing globular cluster in the Galactic bulge from the VVV survey}",
      journal = {\aap},
         year = 2021,
        month = aug,
       volume = {652},
          eid = {A129},
        pages = {A129},
          doi = {10.1051/0004-6361/202140347},
archivePrefix = {arXiv},
       eprint = {2106.13904},
 primaryClass = {astro-ph.GA},
       adsurl = {https://ui.adsabs.harvard.edu/abs/2021A&A...652A.129M}
}

@ARTICLE{miocchi+13,
       author = {{Miocchi}, P. and {Lanzoni}, B. and {Ferraro}, F.~R. and {Dalessandro}, E. and {Vesperini}, E. and {Pasquato}, M. and {Beccari}, G. and {Pallanca}, C. and {Sanna}, N.},
        title = "{Star Count Density Profiles and Structural Parameters of 26 Galactic Globular Clusters}",
      journal = {\apj},
         year = 2013,
        month = sep,
       volume = {774},
       number = {2},
          eid = {151},
        pages = {151},
          doi = {10.1088/0004-637X/774/2/151},
archivePrefix = {arXiv},
       eprint = {1307.6035},
 primaryClass = {astro-ph.GA},
       adsurl = {https://ui.adsabs.harvard.edu/abs/2013ApJ...774..151M}
}

@ARTICLE{montegriffo+95,
       author = {{Montegriffo}, P. and {Ferraro}, F.~R. and {Fusi Pecci}, F. and {Origlia}, L.},
        title = "{IR-array photometry of Galactic globular clusters - II. JK photometry of 47 TUC}",
      journal = {\mnras},
         year = 1995,
        month = oct,
       volume = {276},
       number = {3},
        pages = {739-752},
          doi = {10.1093/mnras/276.3.739},
       adsurl = {https://ui.adsabs.harvard.edu/abs/1995MNRAS.276..739M}
}

@ARTICLE{obasi+21,
       author = {{Obasi}, C. and {G{\'o}mez}, M. and {Minniti}, D. and {Alonso-Garc{\'\i}a}, J.},
        title = "{Confirmation of two new Galactic bulge globular clusters: FSR 19 and FSR 25}",
      journal = {\aap},
         year = 2021,
        month = oct,
       volume = {654},
          eid = {A39},
        pages = {A39},
          doi = {10.1051/0004-6361/202141332},
archivePrefix = {arXiv},
       eprint = {2106.09098},
 primaryClass = {astro-ph.GA},
       adsurl = {https://ui.adsabs.harvard.edu/abs/2021A&A...654A..39O}
}

\end{document}